\documentclass[journal]{IEEEtran}

\ifCLASSINFOpdf
  \usepackage[pdftex]{graphicx}
  \graphicspath{{./}}
  \DeclareGraphicsExtensions{.pdf,.jpeg,.png}
\else
  \usepackage[dvips]{graphicx}
  \graphicspath{{./}}
  \DeclareGraphicsExtensions{.eps,.pdf}
\fi

\usepackage{ifpdf}
\usepackage{cite}
\usepackage[cmex10]{amsmath}
\usepackage{relsize}
\usepackage{algorithmic}
\usepackage{array}
\usepackage{url}
\usepackage{epsfig}
\usepackage{epstopdf}
\usepackage{setspace}
\usepackage{hhline}
\usepackage{times}
\usepackage{amssymb}
\usepackage{etoolbox}
\usepackage{adjustbox}
\usepackage{subcaption}
\usepackage{multirow}
\usepackage{mathtools}
\usepackage{threeparttable}

\DeclareMathSizes{11}{11}{11}{11}
\DeclarePairedDelimiterX{\norm}[1]{\lVert}{\rVert}{#1}

\begin{document}
\title{On Evaluating Perceptual Quality of Online User-Generated Videos}
%
%
%

\author{Soobeom~Jang,
        and~Jong-Seok~Lee,~\IEEEmembership{Senior Member,~IEEE}
\thanks{Manuscript received April 19, 2015; revised February 17, 2016. This research was supported by the MSIP (Ministry of Science, ICT and Future Planning), Korea, under the ``IT Consilience Creative Program'' (IITP-2015-R0346-15-1008) supervised by the IITP (Institute for Information \& Communications Technology Promotion).}%
\thanks{The authors are with the School of Integrated Technology and Yonsei Convergence Institute of Technology, Yonsei University, Incheon, 21983, Korea. Phone: +82-32-749-5846. FAX: +82-32-818-5801. E-mail: soobeom.jang@yonsei.ac.kr;jong-seok.lee@yonsei.ac.kr.}
\thanks{\textcopyright 2016 IEEE. Personal use of this material is permitted. Permission from IEEE must be obtained for all other uses, in any current or future media, including reprinting/republishing this material for advertising or promotional purposes, creating new collective works, for resale or redistribution to servers or lists, or reuse of any copyrighted component of this work in other works.}
\thanks{This is the author's version of an article that has been published in this journal (DOI: 10.1109/TMM.2016.2581582). Changes were made to this version by the publisher prior to publication. The final version of record is available at https://doi.org/10.1109/TMM.2016.2581582.}}

%
%

\markboth{Journal of \LaTeX\ Class Files,~Vol.~13, No.~9, September~2014}%
{Shell \MakeLowercase{\textit{et al.}}: Bare Demo of IEEEtran.cls for Journals}

\maketitle

\begin{abstract}
This paper deals with the issue of the perceptual quality evaluation of user-generated videos shared online, which is an important step toward designing video-sharing services that maximize users' satisfaction in terms of quality. We first analyze viewers' quality perception patterns by applying graph analysis techniques to subjective rating data. We then examine the performance of existing state-of-the-art objective metrics for the quality estimation of user-generated videos. In addition, we investigate the feasibility of metadata accompanied with videos in online video-sharing services for quality estimation. Finally, various issues in the quality assessment of online user-generated videos are discussed, including difficulties and opportunities.
\end{abstract}

\begin{IEEEkeywords}
User-generated video, paired comparison, quality assessment, metadata.
\end{IEEEkeywords}

\IEEEpeerreviewmaketitle

\section{Introduction}
\label{section1}

\IEEEPARstart{W}{ith} the advances of imaging, communications, and internet technologies, public online video-sharing services (e.g., YouTube, Vimeo) have become popular. In such services, a wide range of video content from user-generated amateur videos to professionally produced videos, such as movie trailers and music videos, is uploaded and shared. Today, online video sharing has become the most considerable medium for producing and consuming multimedia content for various purposes, such as fun, information exchange, and promotion.
\par For the success of video-sharing services, it is important to consider users' quality of experience (QoE) regarding shared content, as in many other multimedia services. As the first step of maximizing QoE, it is necessary to measure perceptual quality of the online videos. The quality information of videos can be used for valuable service components such as automatic quality adjustment, streaming quality enhancement, and quality-based video recommendation. The most accurate way to measure perceptual video quality is to conduct subjective quality assessment by employing multiple human subjects. However, subjective quality assessment is not feasible for online videos because of a tremendous amount of videos in online video-sharing services. An alternative is objective quality assessment, which uses a model that mimics the human perceptual mechanism.
\par The traditional objective quality metrics to estimate video quality have two limitations. First, the existence of the reference video, i.e., the pristine video of the given video, is important. In general, objective quality assessment frameworks are classified into three groups: full-reference (FR), reduced-reference (RR), and no-reference (NR). In the cases of FR and RR frameworks, full or partial information about the reference is provided. On the other hand, NR quality assessment does not use any prior information about the reference video, which makes the problem more complicated. In fact, the accuracy of NR objective metrics is usually lower than that of FR and NR metrics \cite{saad2014blind}. Second, the types of degradation that are dealt with are rather limited. Video quality is affected by a large number of factors, for which the human perceptual mechanism varies significantly. Because of this variability, it is too complicated to consider all different video quality factors in a single objective quality metric. Hence, existing objective quality metrics have considered only a single or a small number of major quality factors involved in production and distribution such as compression artifacts, packet loss artifacts, and random noise, assuming that the original video has perfect quality. This approach has been successful for professionally produced videos.
\par However, it is doubtful whether the current state-of-the-art approaches for estimating video quality are also suitable for online videos. First, for a given online video, the corresponding reference video is not available in most cases, where NR assessment is the only option for objective quality evaluation. The performance of existing NR metrics is still unsatisfactory, which makes the quality assessment of online videos very challenging. Second, online video-sharing services cover an extremely wide range of videos. There are two types of videos in online video-sharing services: professional and amateur. Professional videos, which are typically created by professional video makers, and amateur videos, which are created by general users, are significantly different in various aspects such as content and production and editing styles. In particular, user-generated videos have large variations in these characteristics, so they have wide ranges of popularity, user preference, and quality. Moreover, diverse quality factors are involved in online user-generated videos (see Section II for further details). However, existing NR metrics have been developed to work only for certain types of distortion due to compression, transmission error, random noise, etc. Therefore, it is not guaranteed that the existing NR metrics will perform well on those videos.
\par Online videos are usually accompanied with additional information, called metadata, including the title, description, viewcount, rating (e.g., like and dislike), and comments. Some of the metadata of an online video (e.g., the spatial resolution and title) reflect the characteristics of the video signal itself, while other metadata, including the viewcount or comments, provide information about the popularity of and users' preference for the video. These types of information have the potential to be used as hints about the quality of the video because the quality is one of the factors that affects the perceptual preference of viewers. Therefore, they can be useful for the quality assessment of online user-generated videos by replacing or being used in combination with objective quality metrics.
\par This paper deals with the issue of evaluating the perceptual quality of online user-generated videos. The research questions considered are:
\begin{itemize}
\item Are there any noteworthy patterns regarding viewers' judgment of the relative quality of user-generated videos?
\item How well do existing state-of-the-art NR objective quality metrics perform for user-generated videos?
\item To what extent are metadata-based metrics useful for the perceptual quality estimation of user-generated videos?
\item What makes the signal-based or metadata-based quality estimation of user-generated videos difficult?
\end{itemize}
To the best of our knowledge, our work is the first attempt to investigate the issue of the perceptual quality assessment of online user-generated videos comprehensively in various aspects.
Our contributions can be summarized as follows. First, by examining subjective ratings gathered by crowdsourcing for online user-generated videos, we investigate the viewers' patterns of perceptual quality evaluation. Second, we analyze the performance of state-of-the-art NR quality assessment algorithms, metadata-driven features, and their combination in perceptual quality estimation. The study aims the efficacy and limitations of the signal-based and metadata-based methods. Finally, based on the experimental results, various issues in the quality assessment of online user-generated videos are discussed in detail. We comment on the difficulties and limitations of the quality assessment of user-generated videos in general and provide particular examples demonstrating such difficulties and limitations, helping us understand better the nature of the quality assessment of online videos. 
\par The rest of the paper is organized as follows. Section \ref{section2} describes the background of this study, i.e., visual quality assessment, characteristics of online videos, and previous approaches to the quality assessment of online videos. Section \ref{section3} introduces the dataset used in our study, including video data and subjective data. In Section \ref{section5}, patterns of user perception of online videos are examined via graph analysis. Section \ref{section4} presents the results of quality estimation using NR quality assessment algorithms and metadata. Section \ref{section6} discusses issues of the quality assessment of online user-generated videos. Finally, Section \ref{section7} concludes the paper.


\section{Backgrounds}
\label{section2}

\subsection{Visual Quality Assessment}
\label{section2:qa}
The overall QoE of a video service highly depends on the perceptual visual quality of the videos provided by the service. One way to score the quality of videos is to have the videos are evaluated by human subjects, which is called subjective quality assessment. For many practical multimedia applications, quality assessment with human subjects is not applicable due to the cost and real-time operation constraints. To deal with this, research has been conducted to develop automatic algorithms that mimic the human perceptual mechanism, which is called objective quality assessment.
\par Objective quality assessment metrics are classified into three categories: FR, RR, and NR metrics. FR quality assessment uses the entire reference video, which is the original signal without any distortion or quality degradation. Structural similarity (SSIM)\cite{wang2004image}, multi-scale SSIM (MS-SSIM)\cite{wang2003multiscale}, most apparent distortion (MAD)\cite{larson2010most}, and visual information fidelity (VIF)\cite{sheikh2005visual} are well-known FR quality metrics for images, and motion-based video integrity evaluation (MOVIE)\cite{seshadrinathan2010motion} and spatiotemporal MAD (ST-MAD)\cite{vu2011spatiotemporal} are FR metrics for videos. RR metrics do not need the whole reference signal, but use its partial information. Reduced-reference entropic differencing (RRED)\cite{soundararajan2013video} for images and video quality metric (VQM)\cite{pinson2004new} for videos are examples of RR quality metrics.
\par A challenging situation of objective quality assessment is when there is no reference for the given signal being assessed. Estimating quality from only the given image or video itself is hard, since no prior knowledge of the reference can be utilized. Currently available NR metrics include the blind image integrity notator using discrete cosine transform statistics (BLIINDS-II)\cite{saad2012blind}, the blind/referenceless image spatial quality evaluator (BRISQUE)\cite{mittal2012no}, and the Video BLIINDS (V-BLIINDS)\cite{saad2014blind}. These metrics typically use natural scene statistics (NSS) as prior knowledge of images and videos, and the main difference among them lies in how to obtain information about NSS. BLIINDS-II constructs NSS models from the probability distribution of discrete cosine transform (DCT) coefficients extracted from macroblocks. BRISQUE uses mean-subtracted contrast-normalized (MSCN) coefficients rather than transform domain coefficients to speed up the quality assessment process. V-BLIINDS extracts NSS features based on a DCT-based NSS model as in BLIINDS-II, but uses the frame difference to obtain the spatiotemporal information of the video. Additionally, V-BLIINDS uses motion estimation techniques to examine motion consistency in the video.

\subsection{Characteristics of Online Videos}
\label{section2:onlinevideo}

\begin{table*}[!t]
\small
\centering
\caption{\label{table_degradation} Types of observable quality degradation factors in online user-generated videos.}
\resizebox{\textwidth}{!}{
\begin{tabular}{|l||l|l|} \hline
&Step&Types of degradation factors\\ \hline\hline
\multirow{5}{*}{Video}
&\multirow{2}{*}{Acquisition}&Limited spatial/temporal resolution, misfocusing, blur, jerkiness, camera shaking, noise, \\
& &occlusion, insufficient/excessive lighting, poor composition, poor color reproduction \\\cline{2-3}
&\multirow{2}{*}{Processing/editing}&Bad transition effect (e.g., fade in/out, overlap), harming caption (e.g., title screen, subtitle), \\
& &frame-in-frame, inappropriate image processing \\\cline{2-3}
&\multirow{1}{*}{Uploading}&Video compression artifacts, temporal resolution reduction, spatial resolution loss\\\hline
\multirow{4}{*}{Audio}
&Acquisition&Device noise, environmental noise, too-loud or too-low volume, incomprehensible language\\\cline{2-3}
&\multirow{2}{*}{Processing/editing}&Inappropriate background music, audio-video desynchronization, unsuitable sound effects,\\
& & audio track loss\\\cline{2-3}
&Uploading&Audio compression artifacts\\\hline
\multirow{2}{*}{Video \& Audio}
&Content&Boredom, violence, sexuality, harmful content\\\cline{2-3}
&Delivery&Buffering, packet loss, quality fluctuation\\\hline
\end{tabular}}
\end{table*}

\par In online video-sharing services, both user-generated videos and professional videos are shared. In terms of quality, they have significantly different characteristics, especially in filming and editing. Many of the makers of user-generated videos do not have professional knowledge of photography and editing, so quality degradation factors can easily be involved in every step, from the acquisition to the distribution of videos.
\par Table \ref{table_degradation} presents a list of observable quality degradation factors in online user-generated videos. They are grouped with respect to channels affected by degradation (i.e., video, audio, and audio-video) and steps involving degradation.
\par Visual factors in the acquisition step consist of problems with equipment, camera skills, and environments. In particular, according to the work in \cite{wilk2014influence}, typical artifacts in user-generated videos include camera shake, harmful occlusions, and camera misalignment. Visual quality degradation also occurs during video processing and editing due to the editor's lack of knowledge of photography and video making or their intent. For example, scene transition effects, captions, and frame-in-frame effects, where a small image frame is inserted in the main video frame, can be used, which may degrade visual quality. Image processing (e.g., color modification, sharpening) can be applied during editing, which may not be pleasant to viewers. In the uploading step, the system or the uploader may compress the video or modify the spatial and temporal resolutions of the video, which may introduce compression artifacts, visual information loss, or motion jerkiness. 
\par Audio quality degradation can also occur at each step of acquisition, processing and editing, and uploading. Some of the audio quality factors involved in the acquisition and uploading steps are similar to the visual quality factors (equipment noise and compression, etc.). Moreover, the language used in the recorded speech may have a negative effect on perception when a viewer does not understand the language. In the processing and editing step, inserting inappropriate sound sources, background music, or sound effects may decrease user satisfaction. Loss of the audio track may be a critical issue when the content significantly depends on the sound.
\par Some quality factors related to the characteristics of the content or communication environment apply to both audio and video channels. First, the content of a video can be a problem. Boring, violent, sexual, and harmful content can spoil the overall experience of watching the video. Second, the communication environment from the server to a viewer is not always guaranteed, so buffering, packet loss, and quality fluctuation may occur, which are critical in streaming multimedia content \cite{hossfeld2014qoe}.
\par Content in online video-sharing services is usually accompanied with the information reflecting uploading and consumption patterns, provided by uploaders and viewers, which is called metadata. The metadata of a video clip, either assigned by the uploader or automatically extracted from the video, include the title, information of the uploader, upload date, duration, video format, and category. Metadata determined by viewers include the viewcount, comments, and ratings. One can analyze metadata to discover the production and consumption patterns of online videos and to improve the quality of service. Moreover, information from metadata (e.g., video links and subscribers) can be used to construct a social network consisting of online videos. Analysis of the social network can be used for content recommendations \cite{davidson2010youtube} and investigating the network topology of online video sharing, especially the evolution of online video communities \cite{cha2009analyzing} \cite{figueiredo2014dynamics}.

\subsection{Quality Assessment of Online Content}
\label{section2:onlineqa}

There are few studies that consider particular characteristics of online images and videos. The method proposed in \cite{xia2010visual} estimates the quality of online videos using motion estimation, temporal factors to evaluate jerkiness and unstableness, spatial factors (including blockiness, blurriness, dynamic range, and intensity contrast), and video editing styles (including shot length distribution, width, height, and black side ratio). Since these features depend on the genre of video content, robustness is not guaranteed, as pointed out in \cite{xia2010visual}. The work in \cite{yang2014multi} predicted the quality of user-generated images using their social link distribution, the tone of viewer comments, and access logs from other websites. It was discovered that social functionality is more important in determining the quality of user-generated images in online environments than the distortion of the images themselves. Our work deals with videos, which are more challenging for quality assessment than images. In comparison to the aforementioned prior work, we conduct a more comprehensive and thorough analysis of the issue of the quality assessment of online user-generated videos based on state-of-the-art video quality metrics and metadata-driven metrics.


\section{Video and Subjective Dataset}
\label{section3}

\begin{table*}[!t]
\small%
\centering
\caption{\label{table_videos}Description and quality degradation factors of the videos in the dataset. The video ID is defined based on the ranking in terms of the mean opinion score (MOS) (see Section \ref{section5:hodgerank}).}
\centering
\resizebox{\textwidth}{!}{
\begin{tabular}[pos=b]{|l|l|l|} \hline
ID & Description & Quality degradation factors \\\hline \hline
\multirow{2}{*}{1} & A hand drawing portraits on paper with pencil & \multirow{2}{*}{Fast motion} \\
                   & and charcoal from scratch &  \\\hline
2 & A man drawing on the floor with chalk & Time lapse \\\hline
3 & A series of animal couples showing friendship & Blur, compression artifacts \\\hline
4 & Procedure to cook cheese sandwich (close-up of the food) & Misfocusing \\\hline
5 & Two men imitating animals eating food & Compression artifacts, jerkiness \\\hline
6 & A baby swimming in a pool & Camera shaking \\\hline
7 & Escaping from a chase (first-person perspective) & Fisheye lens effect \\\hline
8 & Nature scenes including mountain, cloud, and sea & Blur, compression artifacts \\\hline
\multirow{2}{*}{9} & Cheering university students & \multirow{2}{*}{Camera shaking, captions} \\
                   & (shot by a camera moving around a campus) & \\\hline
10 & A red fox in a cage & Blur, camera shaking \\\hline
11 & Cats and kittens in a house & Blur, misfocusing \\\hline
12 & A crowd dancing at a station & Poor color reproduction \\\hline
13 & Seven people creating rhythmic sounds with a car & Camera shaking, misfocusing \\\hline
14 & People dancing at a square & Camera shaking \\\hline
15 & A group of children learning to cook & Jerkiness, misfocusing \\\hline
16 & A slide show of nature landscape pictures & Blur, compression artifacts \\\hline
17 & Soldiers patrolling streets & Camera shaking, compression artifacts \\\hline
18 & A sleeping baby and a cat (close-up shot) & Camera shaking, compression artifacts \\\hline
19 & A baby smiling at the camera & Poor color reproduction, compression artifacts \\\hline
20 & A man playing a video game & Frame-in-frame, jerkiness \\\hline
21 & Cats having fun with water & Blur, camera shaking, captions \\\hline
22 & A man playing a video game & Frame-in-frame, jerkiness \\\hline
23 & Twin babies talk to each other with gestures & Camera shaking, compression artifacts \\\hline
24 & Walking motion from the walker's viewpoint & Compression artifacts, camera noise \\\hline
25 & A man sitting in a car and singing a song & Compression artifacts, misfocusing \\\hline
26 & A man playing with a pet bear and a dog on the grass & Misfocusing, shaking image frame \\\hline
27 & Pillow fight on a street & Insufficient lighting, varying illumination \\\hline
28 & Kittens playing with each other & Blur, weak compression artifacts \\\hline
29 & People dancing and cheering outside & Compression artifacts, packet loss, excessive lighting \\\hline
30 & A baby laughing loudly & Camera noise, compression artifacts, captions \\\hline
31 & A man breakdancing in a fitness center & Packet loss, compression artifacts, camera noise \\\hline
32 & Cheerleading in a basketball court & Compression artifacts, misfocusing \\\hline
33 & Microlight flying (first-person perspective) & Blur, compression artifacts, misfocusing, packet loss \\\hline
\multirow{2}{*}{34} & \multirow{2}{*}{A man participating in parkour and freerunning} & Compression artifacts, misfocusing, camera shaking, \\
                    & & varying illumination \\\hline
35 & Three people singing in a car & Camera shaking, compression artifacts, blur \\\hline
36 & Street posing performance & Camera shaking, occlusion, blur \\\hline
37 & A puppy and a kitten playing roughly & Low frame rate, compression artifacts, blur, jerkiness \\\hline
38 & Exploring a dormitory building (first-person perspective) & Camera shaking, compression artifacts, misfocusing \\\hline
39 & Shopping people at a supermarket & Compression artifacts, captions, blur \\\hline
\multirow{2}{*}{40} & \multirow{2}{*}{A man working out in a park} & Low frame rate, insufficient lighting, blur, varying illumination, \\
                    & & poor color reproduction \\\hline
41 & People in a hotel lobby & Camera shaking, misfocusing, occlusion \\\hline
42 & Bike trick performance & Compression artifacts, blur, misfocusing, jerkiness \\\hline
43 & A man cooking and eating chicken & Camera shaking \\\hline
44 & A baby walking around with his toy cart at home & Vertical black sides, camera shaking, varying illumination \\\hline
45 & Men eating food & Camera shaking, misfocusing, compression artifacts \\\hline
46 & An old singing performance clip & Camera noise, poor color reproduction, compression artifacts, blur \\\hline
47 & A man doing sandsack training & Compression artifacts, jerkiness, camera noise \\\hline
\multirow{2}{*}{48} & \multirow{2}{*}{A series of short clips} & Poor color reproduction, varying illumination, \\
                    & & compression artifacts, camera noise \\\hline
\multirow{2}{*}{49} & \multirow{2}{*}{Two men practicing martial arts} & Black line running, poor color reproduction, \\
                    & & jerkiness, blur, packet loss \\\hline
\end{tabular}}
\end{table*}

\begin{table*}
\centering
\caption{\label{table_metadata}Range and position information of the metadata of the videos in the dataset.}
\smallskip 
\begin{threeparttable}
\begin{tabular}[pos=b]{|l||r|r|r|r|} \hline
Metadata&Min&Max&Average&Median \\\hline \hline
\textit{Max resolution (height)} & 
144 & 
1080 & 
- & 
480 \\\hline
\textit{Date\tnote{a}} & 
168 & 
3086 & 
1970 & 
2304 \\\hline
\textit{Duration\tnote{b}} & 
27 & 
559 & 
184.3 & 
163 \\\hline
\textit{\#viewcount} & 
29 & 
96372651 & 
6949894 & 
117220 \\\hline
\textit{\#like} & 
0 & 
364131 & 
28750 & 
1458 \\\hline
\textit{\#dislike} & 
0 & 
22426 & 
1231 & 
17 \\\hline
\textit{\#comment} & 
0 & 
49015 & 
5102 & 
260 \\\hline
\textit{Description length} & 
0 & 
3541 & 
279.9 & 
100 \\\hline
\textit{\#subscribe} & 
0 & 
1552053 & 
105682 & 
567 \\\hline
\textit{\#channel viewcount} & 
486 & 
289704644 & 
31341621 & 
892815 \\\hline
\textit{\#channel comment} & 
0 & 
19499 & 
1691 & 
10 \\\hline
\textit{\#channel video} & 
1 & 
5390 & 
225.6 & 
27 \\\hline
\textit{Channel description length} & 
0 & 
911 & 
91 & 
0 \\\hline
\end{tabular}
\begin{tablenotes}
\item[a] Days until the upload date since YouTube was activated (14 Feb. 2005).
\item[b] In seconds.
\end{tablenotes}
\end{threeparttable}
\end{table*}

In this section, we introduce the video and subjective dataset used in this work. We use the dataset presented in \cite{han2014quality}. In \cite{han2014quality}, 50 user-generated amateur videos and their metadata were collected from YouTube via keyword search, and subjective quality ratings for the videos were obtained from a crowdsourcing-based subjective quality assessment experiment. The description and observed quality degradation factors of the videos are presented in Table \ref{table_videos}.
\par Since the metadata collected in \cite{han2014quality} were rather dated and limited, we collected more detailed and recent metadata for the videos by using YouTube Data API. For each video, the following metadata were gathered: the maximum spatial resolution, upload date, video length, video viewcount, video likes, video dislikes, video favorites, video comments, video description length, channel viewcount, channel dislikes, channel favorites, channel comments, channel description length, and number of uploaded videos of the uploader. The collection process for metadata was conducted in March 2014. While collecting recent metadata, we found that one video in the original dataset was deleted, which was excluded from our experiment. Table \ref{table_metadata} presents the metadata statistics of the videos. It can be seen that the dataset covers a wide range of online user-generated videos in various viewpoints of production and consumption characteristics.
\par The subjective ratings were collected based on the paired comparison methodology \cite{lee2014designing} in \cite{han2014quality}. Subjects were recruited from Amazon Mechanical Turk. A web page showing two randomly selected videos from the dataset in a side-by-side manner was used for quality comparison. Subjects were asked to choose a video with better visual quality than the other. Subjects had to play both videos before entering their ratings to prevent cheating and false ratings. In total, 8,471 paired-comparison results were obtained. Each video was shown 332 times, and each pair was matched 6.78 times on average.


\section{Graph-based Subjective Data Analysis}
\label{section5}
The subjective paired comparison data forms an adjacency matrix representing a set of edges of a graph $G = (V,E)$. Here, $V$ is the set of nodes corresponding to the videos, and $E$ is the set of weighted directed edges, where each weight is the winning count where a video is preferred to another one. Therefore, it is possible to apply graph theory to analyze the subjective data, which aims at obtaining further insight into viewers' patterns of quality perception. In this section, two techniques are adopted, HodgeRank analysis and graph clustering.

\subsection{HodgeRank Analysis}
\label{section5:hodgerank}

\begin{figure*}[!t]
\small
\centering
\resizebox{\textwidth}{!}{
\begin{tabular}{cc}
\adjincludegraphics[trim={{.05\width} {.05\height} {.0\width} {0}},clip,width=3.4in]{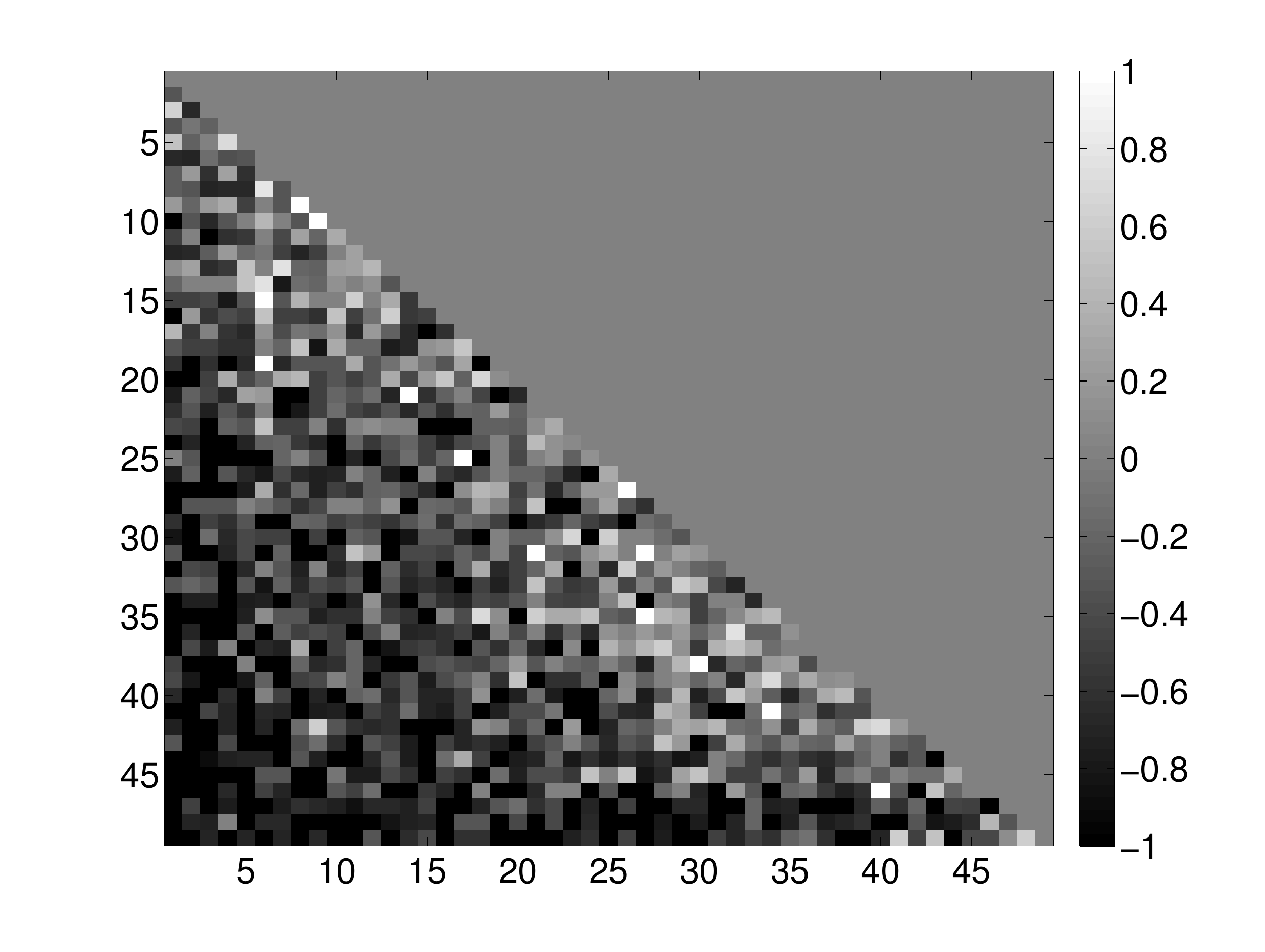}&
\adjincludegraphics[trim={{.05\width} {.05\height} {.0\width} {0}},clip,width=3.4in]{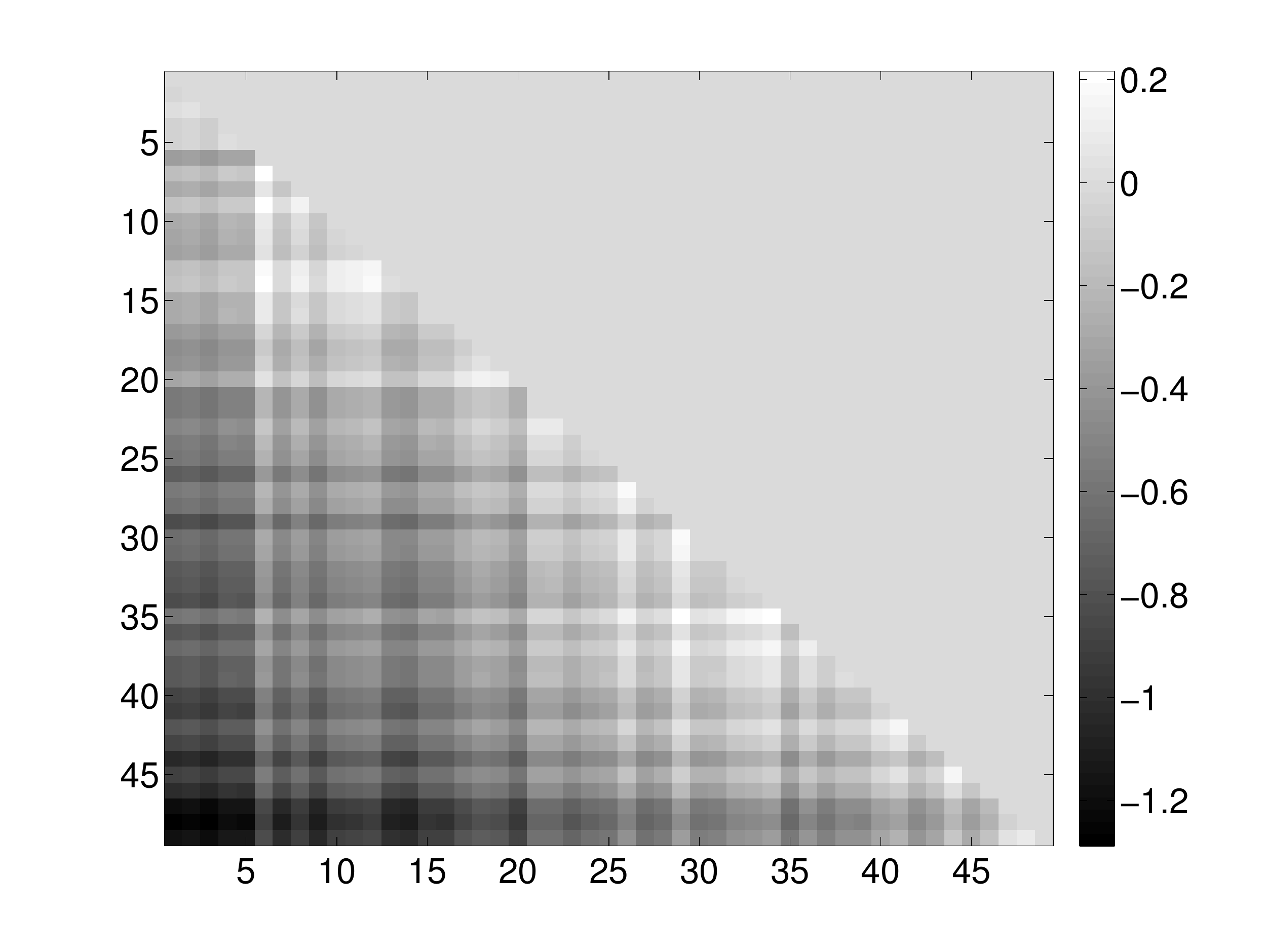}\\
(a) & (b) \\
\adjincludegraphics[trim={{.05\width} {.05\height} {.0\width} {0}},clip,width=3.4in]{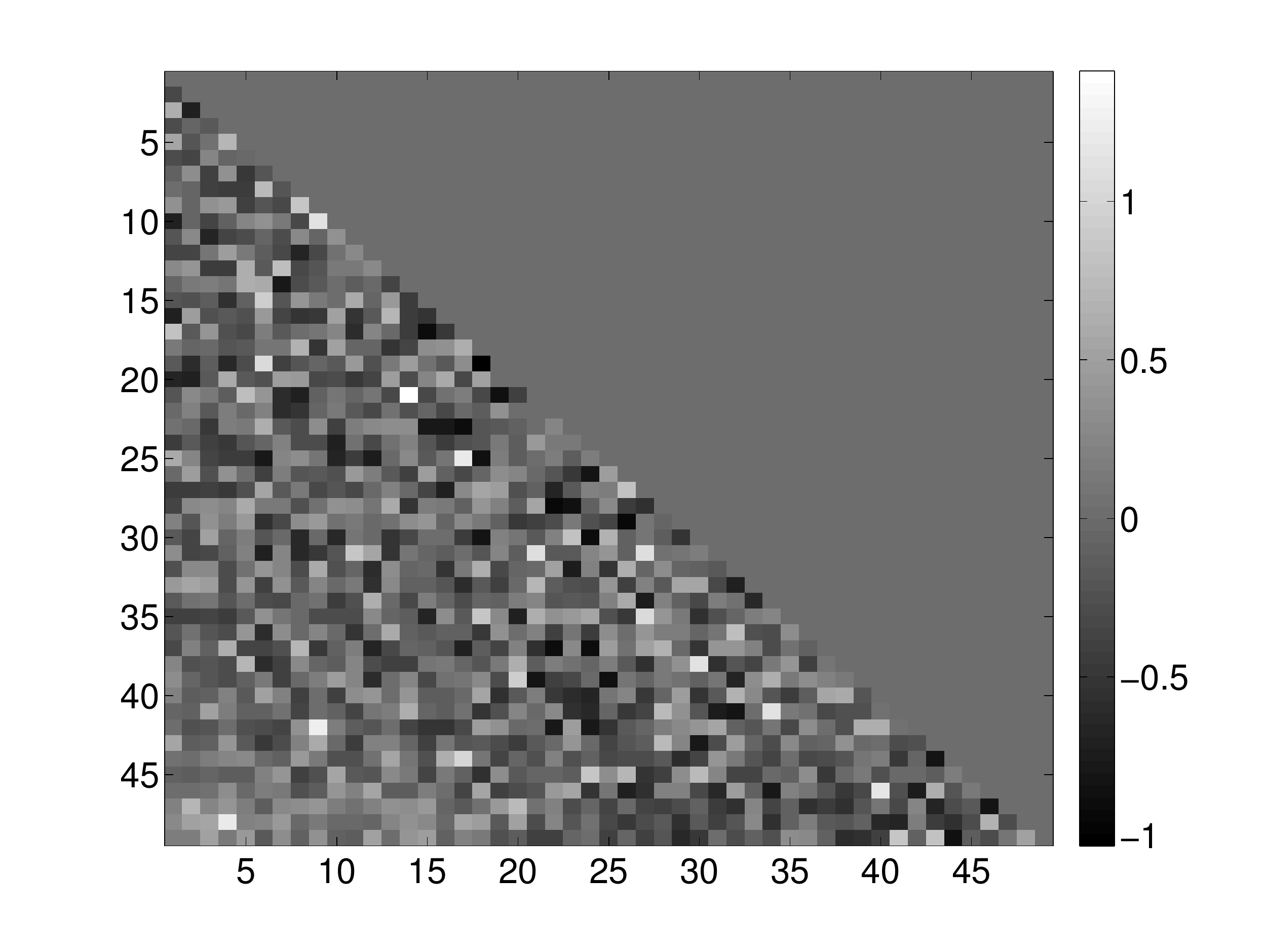}&
\adjincludegraphics[trim={{.05\width} {.05\height} {.0\width} {0}},clip,width=3.4in]{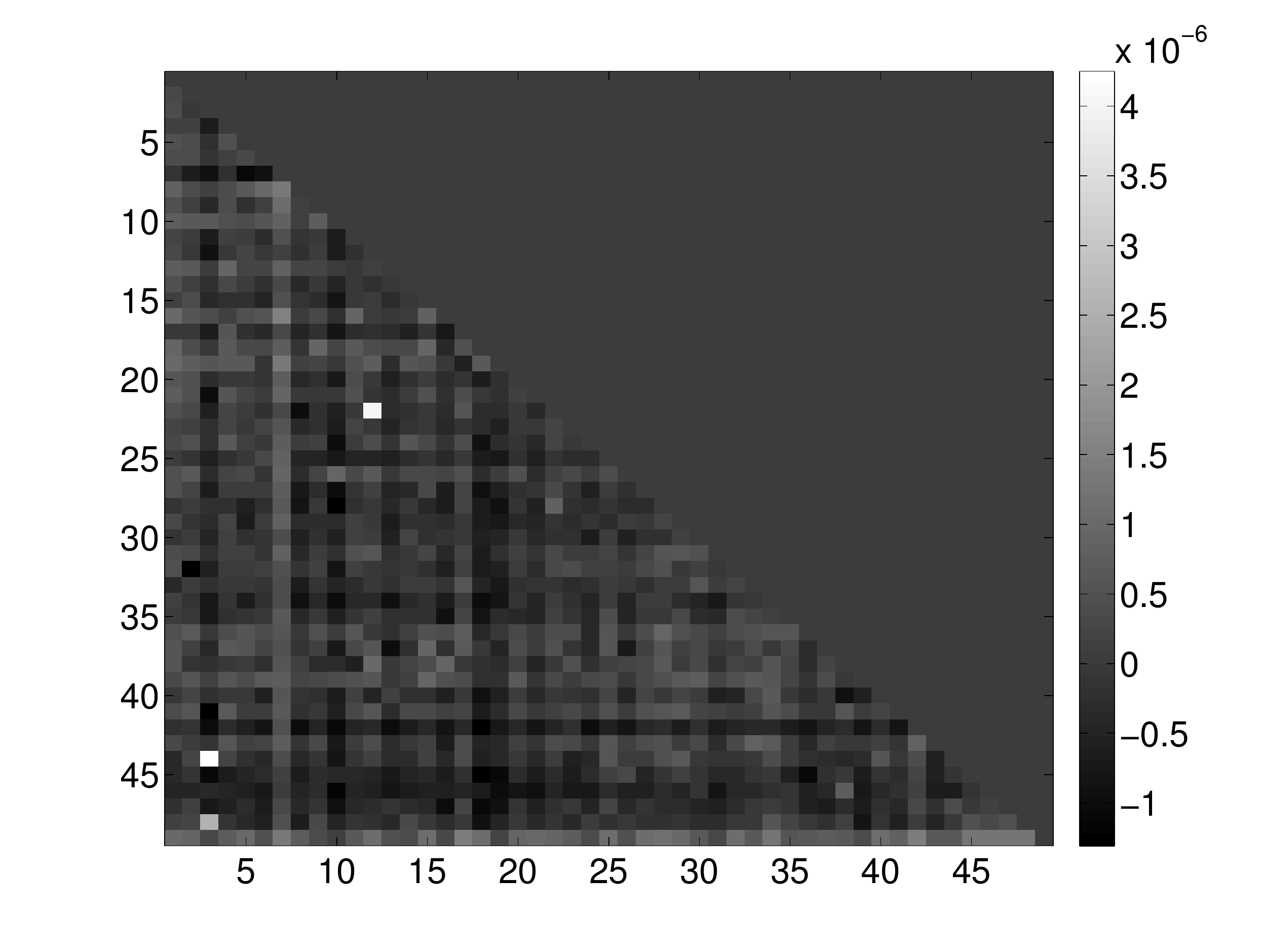}\\
(c) & (d) \\
\end{tabular}}
\caption{\label{fig_hodgerank} Results of HodgeRank analysis. (a) Subjective data matrix $\mathbf{\hat{Y}}$, (b) decomposed global part $\mathbf{\hat{Y}}^g$, (c) decomposed curl part $\mathbf{\hat{Y}}^c$, and (d) decomposed harmonic part $\mathbf{\hat{Y}}^h$. Each axis represents the videos sorted by the global score of HodgeRank. In each figure, only the lower triangular part below the diagonal axis is shown because the matrices are skew-symmetric.}
\end{figure*}

\par The HodgeRank framework introduced in \cite{xu2012hodgerank} decomposes imbalanced and incomplete paired-comparison data into the quality scores of video stimuli and inconsistency of subjects' judgments. In HodgeRank analysis, the statistical rank aggregation problem is posed, which finds the global score $\mathbf{s} = [s_1, s_2, s_3, ..., s_n]$, where $n$ is the number of video stimuli, such that
\begin{center}
\begin{equation}
\label{eq:hodgescore}
\mathlarger{\substack{\mathlarger{\text{min}} \\ \mathbf{s} \in R^n}} \sum_{i,j}w_{ij}(s_i - s_j - \hat{Y}_{ij})^2
\end{equation}
\end{center}
where $w_{ij}$ is the number of comparisons between stimuli $i$ and $j$, and $s_i$ and $s_j$ are the quality scores of stimuli $i$ and $j$, respectively, which are considered as mean opinion scores (MOS). $\hat{Y}_{ij}$ is the $(i,j)$th element of $\mathbf{\hat{Y}}$, which is the subjective data matrix derived from the original graph of paired comparison $G$ by
\begin{center}
\begin{equation}
\label{eq:linearconversion}
\hat{Y}_{ij} = 2\hat{\pi}_{ij}-1
\end{equation}
\end{center}
Here, $\hat{\pi}_{ij}$ is the observed winning rate of stimulus $i$ against stimulus $j$, which is defined as
\begin{center}
\begin{equation}
\label{eq:btmodel}
\hat{\pi}_{ij} = \dfrac{M_{ij}}{M_{ij} + M_{ji}}
\end{equation}
\end{center}
where $M_{ij}$ is the number of counts where stimulus $i$ is preferred to stimulus $j$. Note that $\mathbf{\hat{Y}}$ is skew-symmetric.
\par The converted subjective data matrix $\mathbf{\hat{Y}}$ can be uniquely decomposed into three components as follows, which is called HodgeRank decomposition:
\begin{center}
\begin{equation}
\label{eq:hodgedecomposition}
\mathbf{\hat{Y}} = \mathbf{\hat{Y}}^{g}+\mathbf{\hat{Y}}^{c}+\mathbf{\hat{Y}}^{h}
\end{equation}
\end{center}
where $\mathbf{\hat{Y}}^{g}$,$\mathbf{\hat{Y}}^{c}$, and $\mathbf{\hat{Y}}^{h}$ satisfy the following conditions:

\begin{center}
\begin{equation} \label{eq:hodgecondition1}
\hat{Y}_{ij}^g = \hat{s}_i - \hat{s}_j, 
\end{equation}
\begin{equation} \label{eq:hodgecondition2}
\hat{Y}_{ij}^h + \hat{Y}_{jk}^h + \hat{Y}_{ki}^h = 0, \text{ for } (i,j),(j,k),(k,i) \in E \end{equation}
\begin{equation} \label{eq:hodgecondition2}
\sum_{j \neq i}w_{ij}\hat{Y}_{ij}^h = 0, \text{ for each } i \in V. \end{equation}
\end{center}
where $\hat{s}_{i}$ and $\hat{s}_{j}$ are the estimated scores for stimuli $i$ and $j$, respectively.

\par The global part $\mathbf{\hat{Y}}^{g}$ determines the overall flow of the graph, which is formed by score differences. The curl part $\mathbf{\hat{Y}}^{c}$ indicates the local (triangle) inconsistency (i.e., the situation where stimulus $i$ is preferred to stimulus $j$, stimulus $j$ is preferred to stimulus $k$, and stimulus $k$ is preferred to stimulus $i$ for different $i, j, \text{and } k$). The harmonic part $\mathbf{\hat{Y}}^{h}$ represents the inconsistency caused by cyclic ties involving more than three nodes, which corresponds to the global inconsistency.

\par Fig.\ref{fig_hodgerank} shows the results of the HodgeRank decomposition applied to the subjective data. The overall trend shown in Fig. \ref{fig_hodgerank}(a) is that the total scores decrease (darker color) for elements closer to the lower left corner, showing that the perceived superiority of one video against another becomes clear as their quality difference increases. This is reflected in the global part in Fig. \ref{fig_hodgerank}(b), i.e., the absolute values of the matrix elements increase (darker color).
\par However, Fig. \ref{fig_hodgerank}(a) is also noisy, which corresponds to the inconsistency parts in Fig. \ref{fig_hodgerank}(c) and \ref{fig_hodgerank}(d). To quantify this, we define the ratio of total inconsistency as
\begin{center}
\begin{equation}
\label{eq:totalincon}
\text{Total Inconsistency} = \dfrac{\norm{\mathbf{\hat{Y}}^{h}}_F^2}{\norm{\mathbf{\hat{Y}}}_F^2} + \dfrac{\norm{\mathbf{\hat{Y}}^{c}}_F^2}{\norm{\mathbf{\hat{Y}}}_F^2}
\end{equation}
\end{center}
where $\norm{\cdot}_F$ is the Frobenius norm of a matrix. The obtained ratio of total inconsistency for the subjective data is 67\%. That is, the amount of inconsistency, including local inconsistency and global inconsistency, is larger than that of the global flow of the graph. Between the two sources of inconsistency, the amount of the harmonic component is far smaller than that of the curl component, as can be seen from the scale of the color bar in Fig. \ref{fig_hodgerank}(d). This implies that it is easy for human subjects to determine quality superiority between videos with significantly different ranks (i.e., quality scores), while determining preference for videos where quality is ranked similarly is relatively difficult. This will be discussed further in Section VI.

\subsection{Graph Clustering}
\label{section5:clustering}

The HodgeRank analysis showed that videos with similar ranks in the MOS are subjectively ambiguous in terms of quality. Therefore, one may hypothesize that the videos can be grouped in such a way that different groups have distinguishable quality differences, while videos in each group have a similar quality. We attempt to examine if this is the case and, if so, how many groups can be found via graph clustering.
\par We use the algorithm presented in \cite{duan2009community}, whose objective is to divide the whole graph represented by the adjacency matrix $\mathbf{M}$ into $l$ groups, $U_1, U_2, ..., U_l$, by maximizing the modularity measure $Q$:

\begin{center}
\begin{equation}
Q = \dfrac{1}{A}\sum_{p=1}^{l} \sum_{i,j \in U_p} \left(M_{ij} - \dfrac{d_{i}^{in}d_{j}^{out}}{A} \right)
\end{equation}
\end{center} 
where $A = \sum_{i,j}M_{ij}$ is the sum of all edge weights in the graph, and $d_{i}^{in} = \sum_{j}M_{ij}$ and $d_i^{out} = \sum_{j}M_{ji}$ represent the in-degree and out-degree of the $i$-th node, respectively.
\par This algorithm is based on the random walk with restart (RWR) model. It first computes the relevance matrix $\mathbf{R}$, which is the estimated result matrix of the RWR model, from the transition matrix, which equals $\mathbf{M}$:
\begin{center}
\begin{equation}
\label{eq:relevance}
\mathbf{R} = (1-\delta)(\mathbf{I}-\delta\mathbf{\tilde{M}})
\end{equation}
\end{center}
where $\delta$ is the restart probability (an adjustable parameter in the RWR model), $\mathbf{I}$ is an identity matrix, and $\mathbf{\tilde{M}}$ is the column-normalized version of $\mathbf{M}$. The algorithm then consists of two steps. First, starting with an arbitrary node, it repeatedly adds the node with the largest single compactness measure, until its single compactness measure does not increase. The single compactness of node $i$ with respect to local cluster $G'$ is represented by:

\begin{figure}[!t]
\begin{center}
\adjincludegraphics[trim={{0} {0} {.05\width} {.05\height}},clip,width=3.5in]{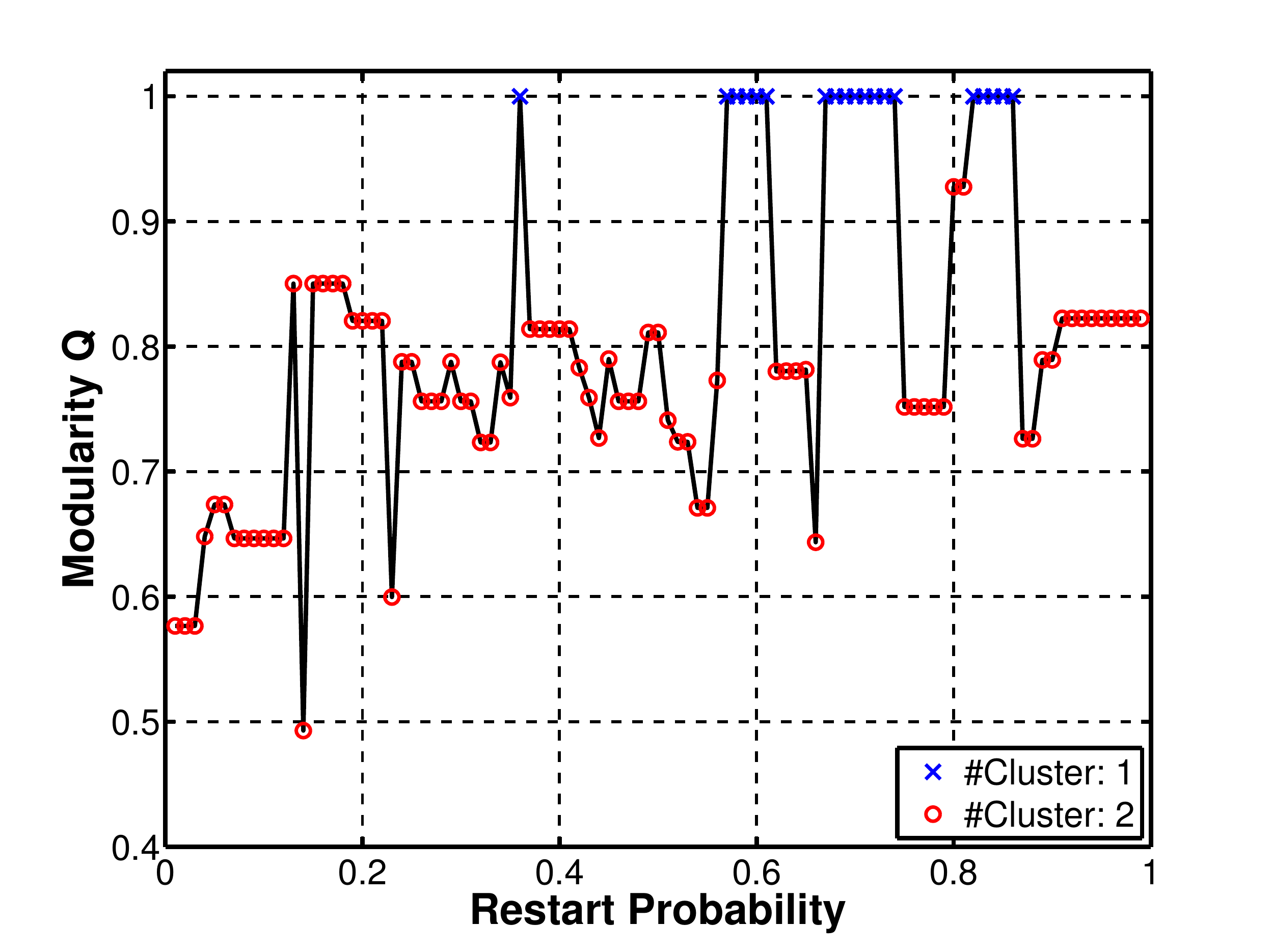}
\caption{Graph clustering results with respect to restart probability $\delta$.}
\label{fig_modularity}
\end{center}
\end{figure}

\begin{center}
\begin{equation}
C(i,G') = \dfrac{1}{B}\left(R_{ii} + \sum_{j \in G'}(R_{ij}+R_{ji}) - \dfrac{R_{i*}R_{*i}}{B} \right)
\end{equation}.
\end{center}
where $B=\sum_{i,j}R_{ij}$ is the total sum of the elements of $\mathbf{R}$, and $R_{i*}=\sum_{k}R_{ik}$ and $R_{*i}=\sum_{k}R_{ki}$ are the row and column sums of $\mathbf{R}$, respectively. If the construction of a local cluster from a node is finished, the algorithm starts from another node that has not been assigned to any local clusters yet to make another local cluster. This process is repeated until all nodes are assigned to certain local clusters. After constructing local clusters, the algorithm merges the compact clusters by maximizing the increase of the total modularity $Q$ of clusters in a greedy manner until there is no increase of $Q$, which results in final clusters.
\par
\begin{figure*}[!t]
\small
\centering
\resizebox{\textwidth}{!}{
\begin{tabular}{ccccc}
\adjincludegraphics[trim={{0} {0} {0} {0}},clip,width=2in]{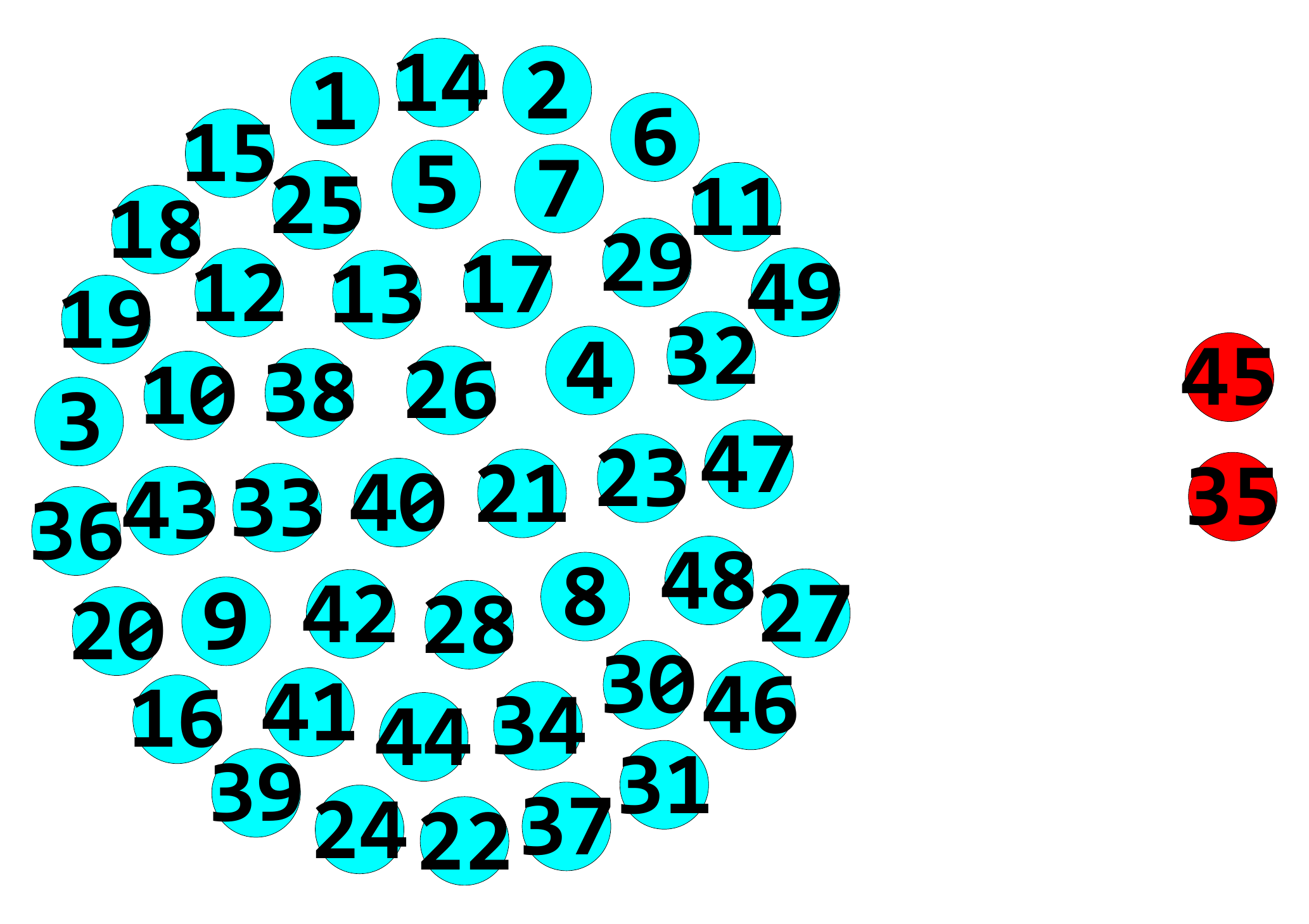}&\vline&
\adjincludegraphics[trim={{0} {0} {0} {0}},clip,width=2in]{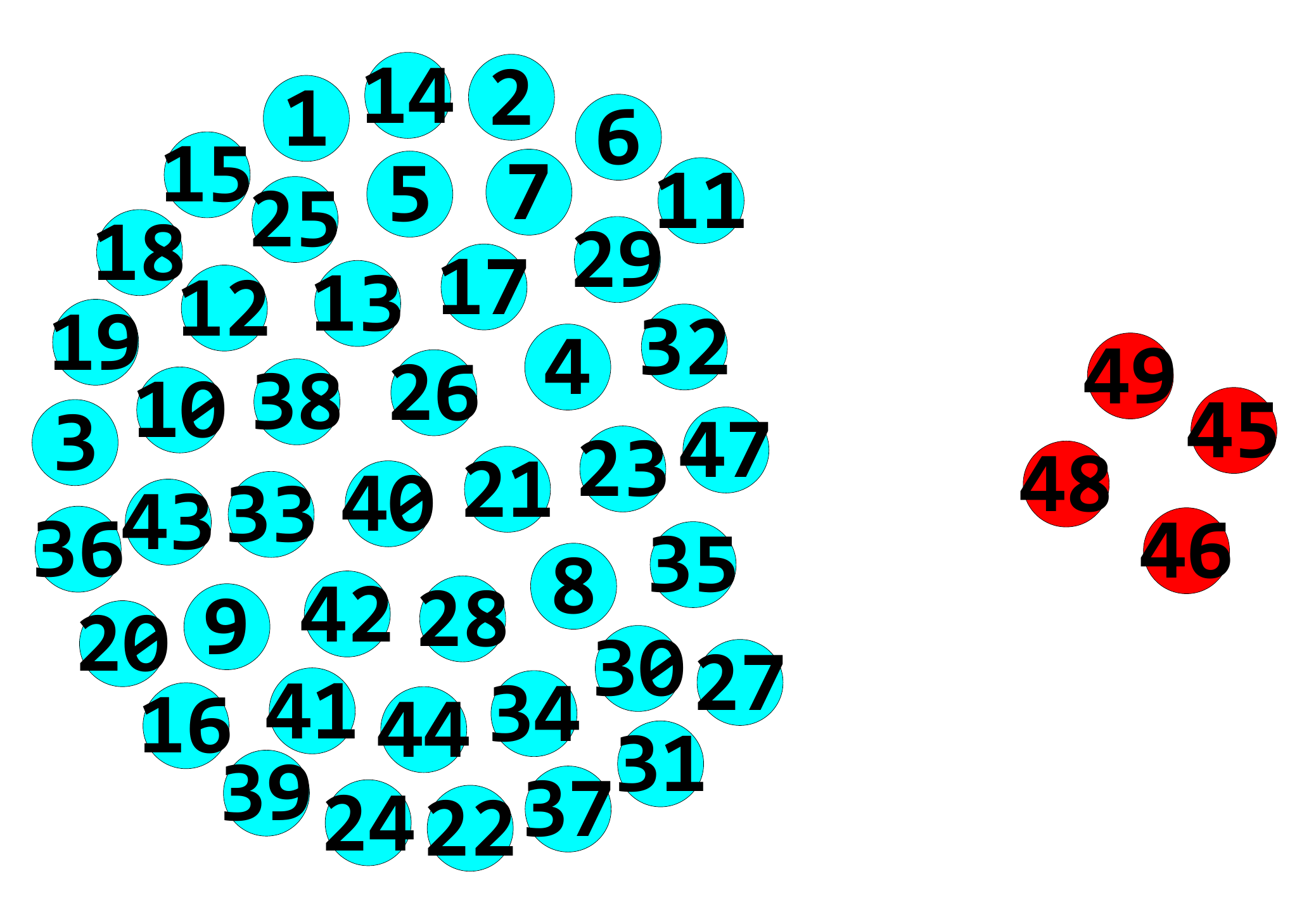}&\vline&
\adjincludegraphics[trim={{0} {0} {0} {0}},clip,width=2in]{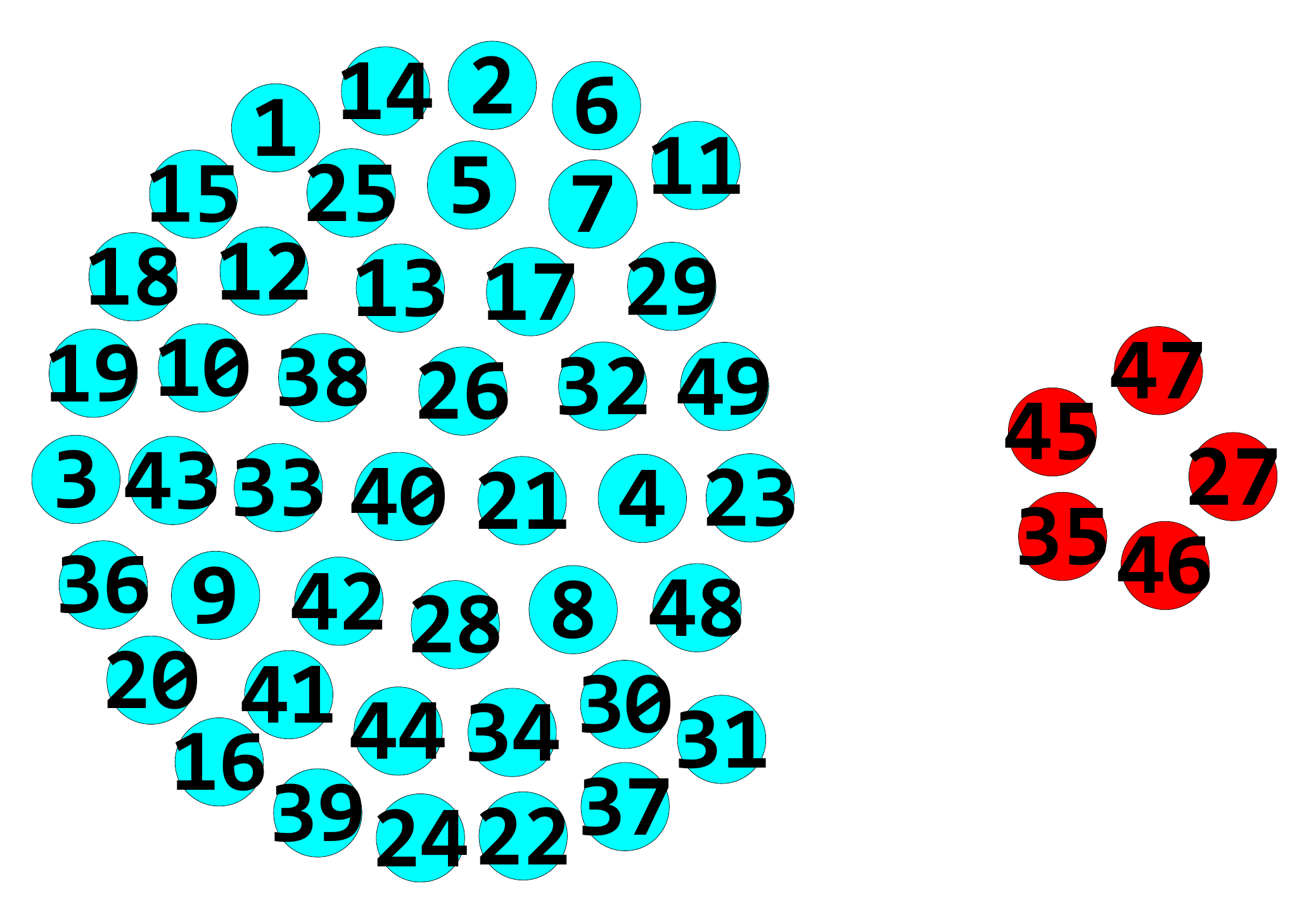}\\
(a) && (b) && (c) \\
\end{tabular}}
\caption{\label{fig_cluster} Examples of graph clustering results. Each video is represented by its ID, and the high (low)-quality group is marked with blue (red). (a) $\delta = 0.81$ ($Q = 0.9275$) (b) $\delta = 0.15$ ($Q = 0.8503$) (c) $\delta = 0.99$ ($Q = 0.8225$)}
\end{figure*}

 We apply the aforementioned algorithm to the subjective data graph for various restart probability values. Fig. \ref{fig_modularity} shows the final modularity value with respect to the restart probability, ranging from 0.01 to 0.99. The number of final clusters differs when the restart probability differs. It can be seen that the graph is clustered into one or two groups in all cases. In particular, the results with $Q=1$ correspond to the case in which all nodes in the graph are assigned to one cluster. That is, it is difficult to divide the nodes into groups with a clearly distinguished subjective preference. Fig. \ref{fig_cluster} shows examples of final clustering results that have high modularity. Two clusters are formed in these examples, and the cluster containing high-quality videos (marked with blue) is much bigger than that containing low-quality videos (marked with red). It seems that in the used video dataset, discriminating quality for videos with high and medium quality was difficult, whereas videos with medium and low quality were more easily distinguished.

\section{Quality Estimation}
\label{section4}

In this section, we investigate the problem of the objective quality assessment of online videos. First, the performance of the state-of-the-art objective quality metrics is evaluated. Second, quality estimation using metadata-driven metrics is investigated.

\subsection{No-Reference Objective Metrics}
\label{section4:noref}
The performance of three state-of-the-art NR quality metrics, namely, V-BLIINDS, BRISQUE, and BLIINDS-II, is examined using the video and subjective data described in the previous section. Some videos have title scenes at the beginning (e.g., title text shown on a black background for a few seconds), which are excluded for quality evaluation because they are usually irrelevant for judging video quality.
\begin{table}[!b]
\small%
\centering
\caption{\label{table_nrvqa} Spearman rank-order correlation coefficients (SROCC) between MOS and V-BLIINDS, BRISQUE, and BLIINDS-II, respectively.}
\centering
\begin{tabular}[pos=b]{|c|c|} \hline
NR Metric&SROCC \\\hline \hline
\textit{V-BLIINDS} & 0.1406\\\hline
\textit{BRISQUE} & 0.3364\\\hline
\textit{BLIINDS-II} & 0.2383\\\hline
\end{tabular}
\end{table}
\par The performance of the metrics is shown in Table \ref{table_nrvqa}. We adopt the Spearman rank-order correlation coefficient (SROCC) as the performance index because the relationship between the metrics' outputs and MOS is nonlinear. Statistical test results reveal that only the SROCC of BRISQUE is statistically significant at a significance level of $0.05$ ($F = 6.00$). Interestingly, V-BLIINDS, which is a video quality metric, appears to be inferior to BRISQUE and BLIINDS-II, which are image quality metrics, meaning that the way V-BLIINDS incorporates temporal information is not very effective for the user-generated videos. Overall, the performance of the metrics is far inferior to that shown in existing studies using other databases. The SROCC values of BLIINDS-II and BRISQUE on the LIVE IQA database containing images corrupted by blurring, compression, random noise, etc. \cite{sheikh2006statistical} were as high as 0.9250 and 0.9395, respectively \cite{gu2015using}. The SROCC of V-BLIINDS for the LIVE VQA database containing videos degraded by compression, packet loss, etc. \cite{seshadrinathan2010study} was 0.759 in \cite{saad2014blind}. This implies that the problems of online video quality assessment are very different from those of traditional quality assessment. The reasons why the NR metrics fail are discussed further in Section \ref{section6} with examples.

\subsection{Metadata-driven Metrics}
\label{section4:metadata}

\begin{table}[!t]
\small%
\centering
\caption{\label{table_metasrocc}SROCC between MOS and metadata-driven metrics. Note that the metrics are sorted in descending order of SROCC.}
\centering
\begin{tabular}[pos=b]{|l||r|} \hline
Metric&SROCC \\\hline \hline
\textit{Description length} & 0.5402 \\\hline
\textit{\#like / \#view} & 0.5347 \\\hline
\textit{Max resolution} & 0.5331 \\\hline
\textit{\#subscribe} & 0.4694 \\\hline
\textit{\#subscribe / \#channel video} & 0.4408 \\\hline
\textit{\#like} & 0.4307 \\\hline
\textit{\#dislike} & 0.3910 \\\hline
\textit{\#channel viewcount} & 0.3855 \\\hline
\textit{\#viewcount} & 0.3728 \\\hline
\textit{\#comment} & 0.3699 \\\hline
\textit{\#like / date} & 0.3558 \\\hline
\textit{Channel description length} & 0.3482 \\\hline
\textit{\#channel viewcount / \#channel video} & 0.3061 \\\hline
\textit{Date} & 0.3042 \\\hline
\textit{\#view / date} & 0.2727 \\\hline
\textit{\#channel comment} & 0.2099 \\\hline
\textit{\#comment / \#view} & 0.1861 \\\hline
\textit{\#channel video} & 0.1465 \\\hline
\textit{\#channel comment / \#channel video} & 0.1414 \\\hline
\textit{Duration} & -0.0533 \\\hline
\end{tabular}
\end{table}

\par Metadata-driven metrics are defined as either the original values of the metadata listed in Table \ref{table_metadata} or the values obtained by combining them (e.g., \#like divided by \#view for normalization). Table \ref{table_metasrocc} shows the performance of the metadata-driven metrics for quality prediction. It is observed that the performance of the metrics significantly varies, from fairly high to almost no correlation with the MOS. It is worth noting that several metadata-driven metrics show better performance than the NR quality metrics. Generally, the video-specific metrics (e.g., video viewcount, the number of video comments) show better performance than the channel-specific metrics (e.g., channel viewcount, number of channel comments).

\begin{figure*}[!t]
\small
\centering
\resizebox{\textwidth}{!}{
\begin{tabular}{cc}
\adjincludegraphics[trim={{0} {0} {.05\width} {.05\height}},clip, width=3.4in]{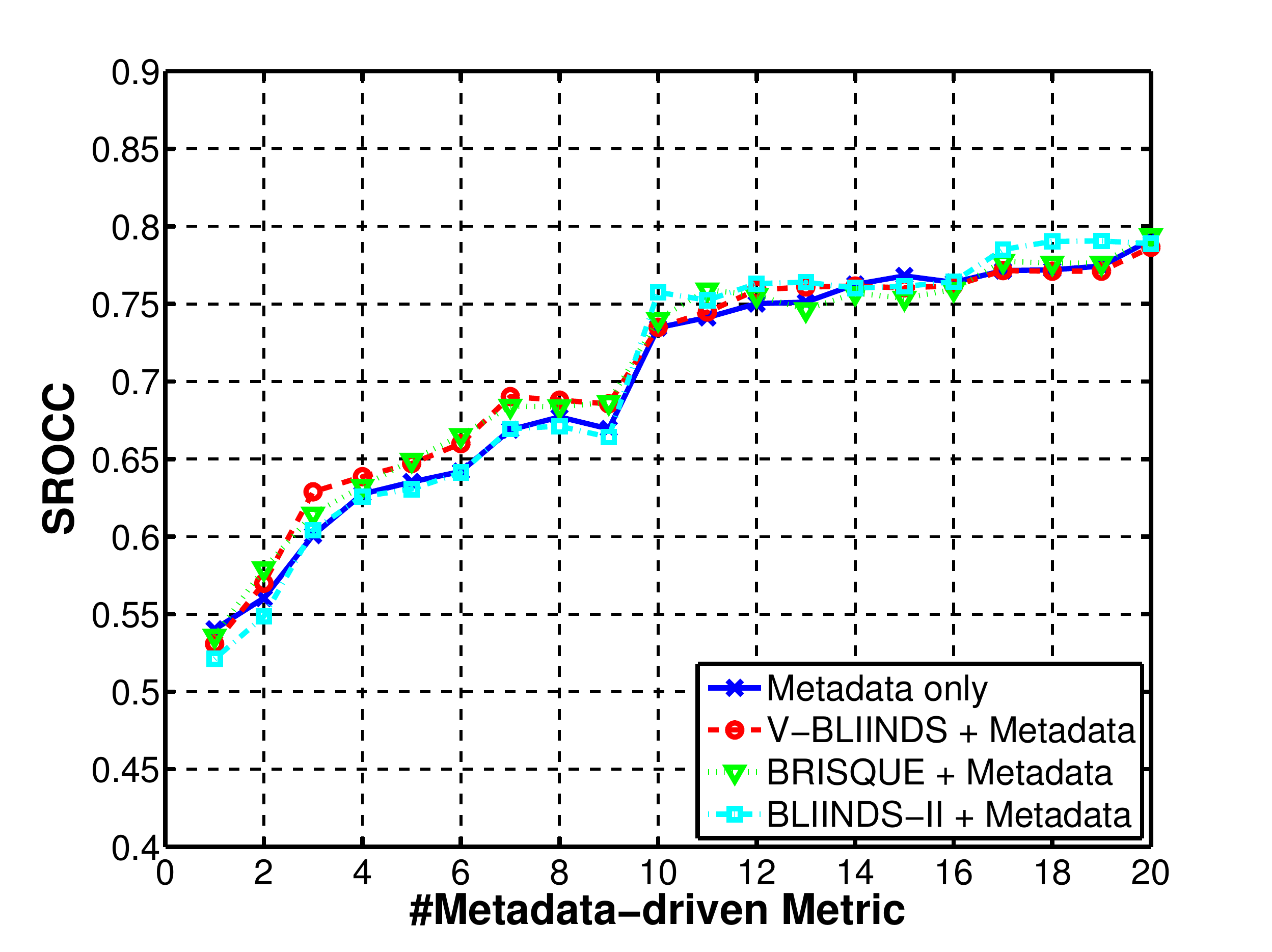}&
\adjincludegraphics[trim={{0} {0} {.05\width} {.05\height}}, clip, width=3.4in]{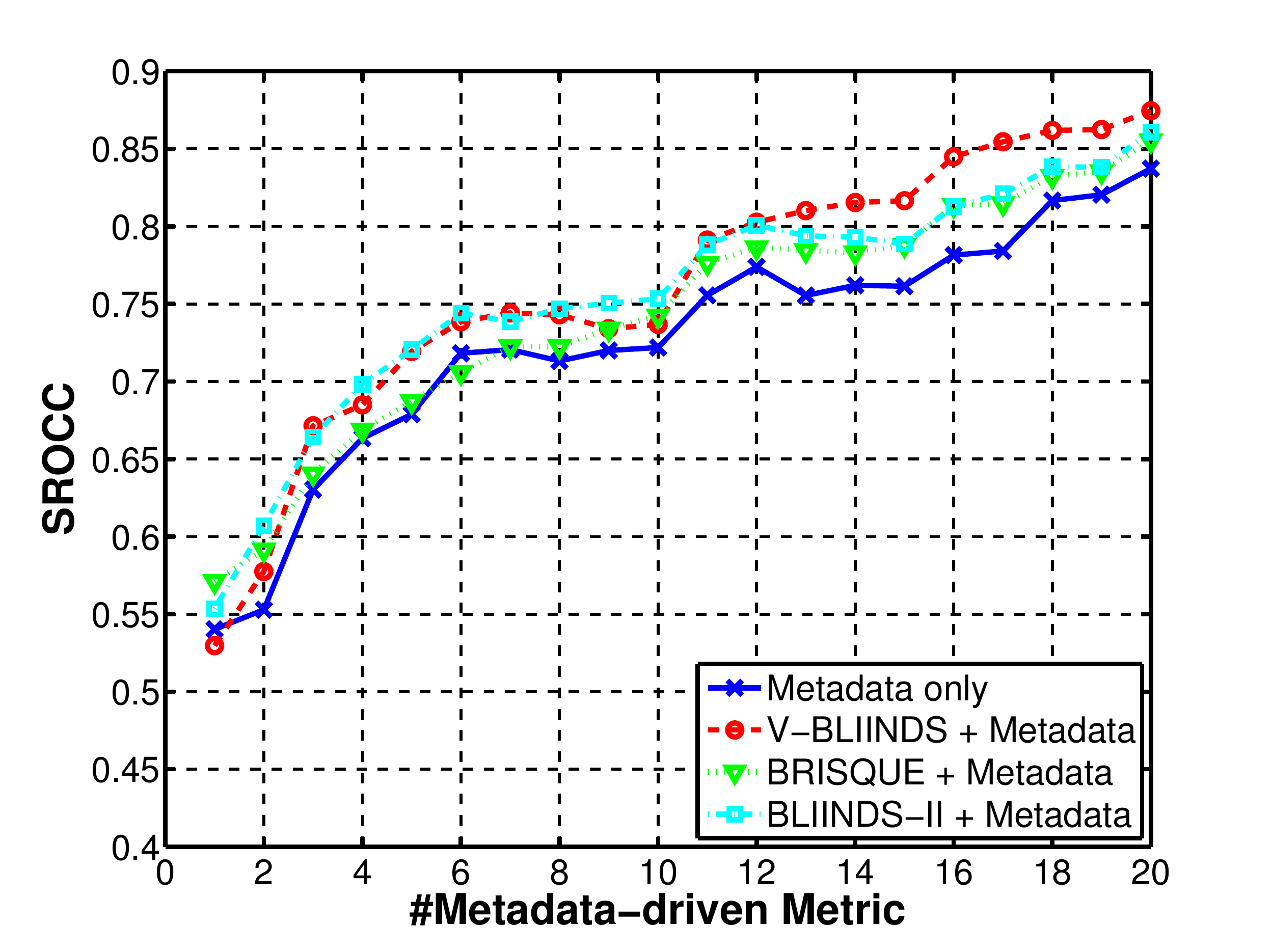}\\
\hspace{0.2in}(a) & \hspace{0.2in}(b)
\end{tabular}}
\caption{\label{fig_linearsvr}SROCC of (a) linear regression and (b) SVR using metadata and objective quality assessment algorithms.}
\end{figure*}

\par The metadata-driven metric showing the highest SROCC is the description length. A possible explanation for this is that a video with good quality would have significant visual and semantic information, and the uploader may want to provide a detailed description about the video faithfully. The second-best metadata-driven metric is the ratio of the like count to the viewcount. This metric inherently contains information about the satisfaction of the viewers, which is closely related to the video quality. The third-ranked metadata-driven metric is the maximum spatial resolution. The availability of high resolution for a video means that it was captured with a high-performance camera or that it did not undergo video processing, which would reduce the spatial resolution and possibly degrade video quality. The number of subscribers of an uploader is ranked fourth, which indicates the popularity of the uploader. A popular uploader's videos are also popular, and their visual quality plays an important role. 
\par Other metrics related to video popularity, including the numbers of likes and dislikes, viewcount, and number of comments show significant correlations with the MOS, but they remain moderate. This indicates that popularity is a meaningful but not perfect predictor of quality. The age of a video (``date'' in the table) has a moderate correlation with perceptual quality, which means that newer videos tend to have better quality. This can be understood by considering that recent recording devices usually produce better quality videos than old devices and users are more experienced in video production than before.
\par Apparently, each of the metadata-driven metrics represents a distinct partial characteristic of the videos. Therefore, it can be expected that combining multiple metrics will yield improved results due to their complementarity. For the integration of the metrics, we employ two techniques, namely, linear regression and nonlinear support vector regression (SVR) using Gaussian kernels \cite{smola1997support}. The former is written as:
\begin{center}
\begin{equation}
\label{eq:linear}
f_{Linear}(\mathbf{x}) = \mathbf{a}^T\mathbf{x} + b_{Linear}
\end{equation}
\end{center}
where $\mathbf{x}$ is a vector composed of metadata-driven metrics, and $\mathbf{a}$ and $b_{Linear}$ are tunable parameters of the linear regression model. The latter is given as:
\begin{center}
\begin{equation}
\label{eq:svr}
f_{SVR}(\mathbf{x}) = \sum_{i=1}^{v}(\alpha_i - \alpha_i^{*})\phi(\mathbf{x}_i,\mathbf{x})+b_{SVR}
\end{equation}
\end{center}
where $\alpha_i$, $\alpha_i^{*}$, $\mathbf{x}_i$ ($i=1,2,...,v$), and $b_{SVR}$ are parameters of the SVR model, and $\phi(\mathbf{x}_i, \mathbf{x})$ is a Gaussian kernel expressed by
\begin{center}
\begin{equation}
\phi(\mathbf{x}_i, \mathbf{x}) = \text{exp}(-\dfrac{\norm{\mathbf{x}_i - \mathbf{x}}^2}{2\sigma^2})
\end{equation}
\end{center}
where $\sigma^2$ is the variance of the Gaussian kernel \cite{smola2004tutorial}.

\begin{figure*}[!t]
\small
\centering
\resizebox{\textwidth}{!}{
\begin{tabular}{cc}
\adjincludegraphics[trim={{0} {0} {.05\width} {0}},clip,width=3.4in]{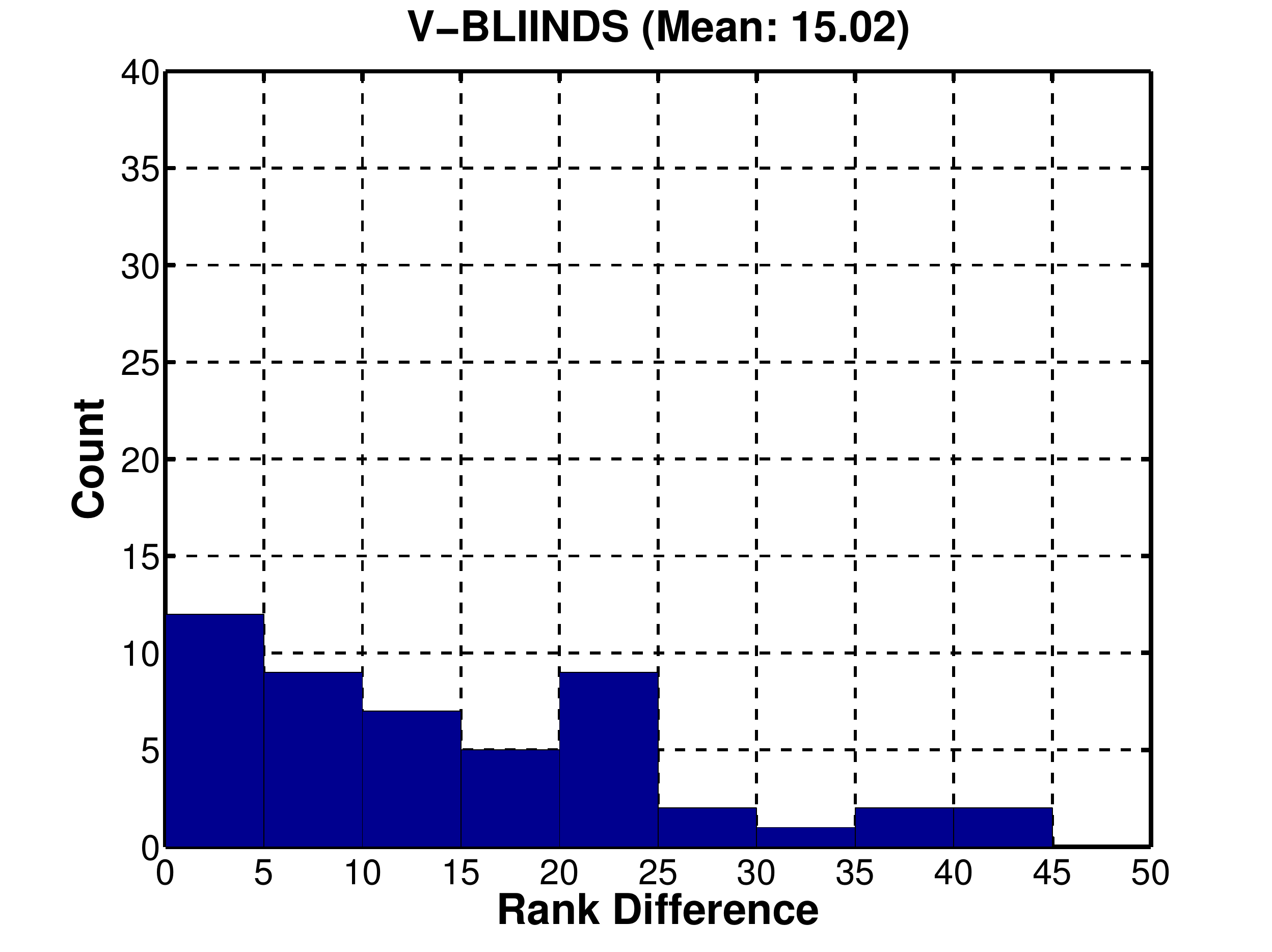}&
\adjincludegraphics[trim={{0} {0} {.05\width} {0}},clip,width=3.4in]{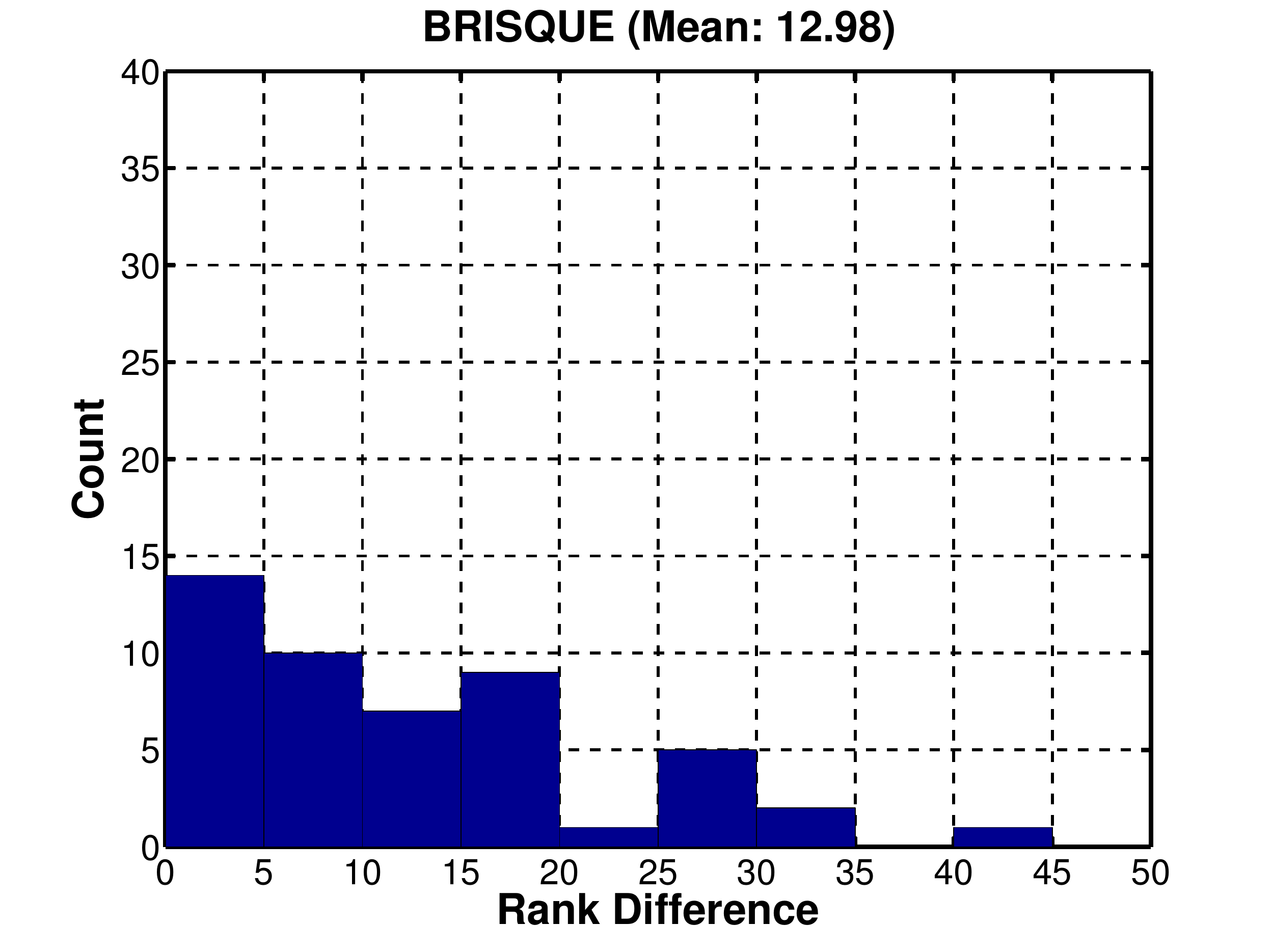}\\
\hspace{0.2in}(a) & \hspace{0.2in}(b) \\
\adjincludegraphics[trim={{0} {0} {.05\width} {0}},clip,width=3.4in]{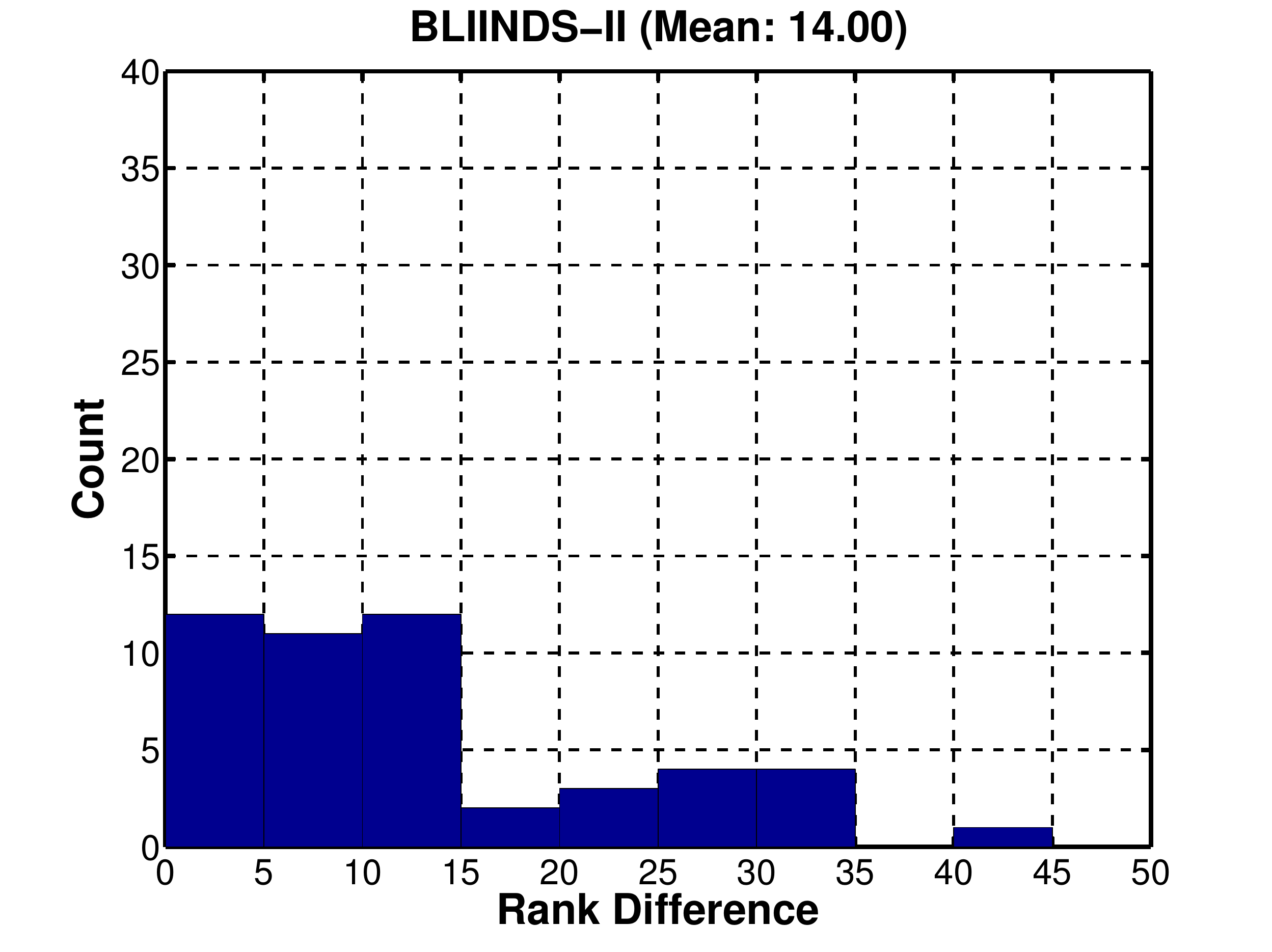}&
\adjincludegraphics[trim={{0} {0} {.05\width} {0}},clip,width=3.4in]{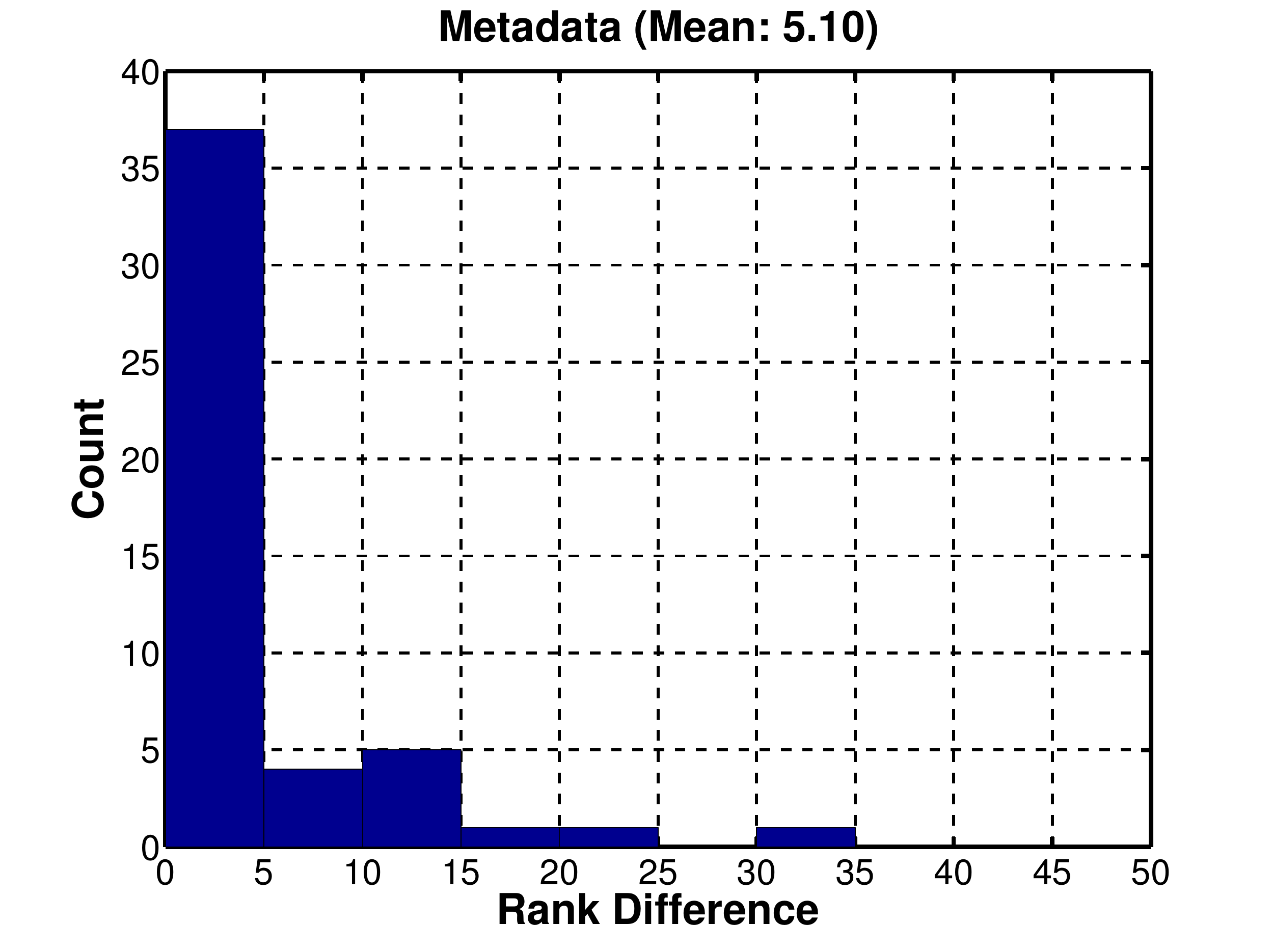}\\
\hspace{0.2in}(c) & \hspace{0.2in}(d)
\end{tabular}}
\caption{\label{fig_rankdiff} Rank difference between estimated quality scores and MOS. (a) V-BLIINDS, (b) BRISQUE, (c) BLIINDS-II, and (d) the SVR model with 20 metadata-driven metrics.}
\end{figure*}

\par We train the models by changing the number of metadata-driven metrics, i.e., the length of the input vector $\mathbf{x}$ in (\ref{eq:linear}) and (\ref{eq:svr}), by adding metrics one by one in descending order of SROCC in Table \ref{table_metasrocc}. The models are trained and evaluated using all the data without separating the training and test data to obtain insight into the performance potential of the integrated models to estimate quality. In addition, to further incorporate the information of visual signals within the regression models, we also test models combining metadata-driven metrics and V-BLIINDS, BRISQUE, and BLIINDS-II, respectively, to examine the synergy between the two modalities (i.e., the visual signal and metadata).
\par Fig. \ref{fig_linearsvr} shows the results of the linear regression and SVR model evaluation. In both cases, the performance in terms of SROCC is improved by increasing the numbers of the metadata-driven metrics. When 20 metadata-driven metrics are used, the SROCC becomes nearly 0.8 for linear regression and 0.85 for SVR, which is a significant improvement compared with the highest SROCC achieved by a single metric (0.54 by the description length). The effect of incorporating an NR objective quality metric is negligible (for linear regression) or marginal (for SVR). Such a limited contribution of the video signal-based metrics seems to be due to their limited quality predictability as shown in Table \ref{table_nrvqa}.
\par To compare the quality prediction performance of metadata and NR quality metrics in more depth, we analyze differences between the predicted ranks and MOS ranks. Fig. \ref{fig_rankdiff} shows histograms of rank differences between the MOS and estimated quality scores by the NR objective quality metrics or SVR combining 20 metadata-driven metrics. It is observed that the SVR model of metadata shows smaller rank differences than the objective quality metrics. From an ANOVA test, it is confirmed that the mean locations of the rank differences for the four cases are significantly different at a significance level of $0.01$ ($F = 10.84$). Additionally, Duncan's multiple range tests reveal that the mean location of the rank differences of the SVR-based metadata model is significantly different from those of the signal-based metrics at a significance level of $0.01$. These results demonstrate that metadata-based quality prediction is a powerful way of dealing with the online video quality assessment problem by overcoming the limitations of the conventional objective metrics based on the visual signal.


\section{Discussion}
\label{section6}
In the previous sections, it was shown that the quality evaluation of online user-generated videos is not easy both subjectively and objectively. In this section, we provide further discussion on these issues in more detail with representative examples.
\par In Section \ref{section5}, it was observed that viewers have difficulty in determining quality superiority among certain videos. As shown in Table \ref{table_degradation}, there are many different factors of quality degradation in user-generated videos. In many cases, therefore, users are required to compare quality across different factors, which is not only difficult, but also very subjective depending on personal taste. For example, video \#15 has jerkiness and misfocusing, video \#16 has blur, and video \#17 has camera shaking. In the subjective test results, video \#16 is preferred to video \#15, while video \#17 is less preferred than video \#15, and video \#17 is preferred to video \#16. As a result, the match result among these three videos forms a triangular triad, which contributes to the local inconsistency shown in Fig. \ref{fig_hodgerank}(c). 
\par In Section \ref{section4}, we showed that the state-of-the-art NR objective metrics fail to predict the perceptual quality of user-generated videos. This is largely due to the fact that the quality degradation conditions targeted during the development of the metrics are significantly different from those involved in user-generated videos. When the NR metrics were developed, it was normally assumed that original video sequences have perfect quality. However, as shown in Table \ref{table_degradation}, the user-generated videos are already subject to various types of quality degradation in the production stage (e.g., insufficient lighting, hand shaking). Furthermore, NR metrics are usually only optimized for a limited number of typical quality factors, such as compression artifacts, packet loss, blurring, and random noise, while there are many more quality factors that can be involved during editing, processing, and distribution, some of which are even related to aesthetic aspects. 
\par Editing effects are particularly difficult to assess for the NR metrics, not only because many of them are not considered by the metrics, but also some of them may be wrongly treated as artifacts. Videos \#1 and \#2 are examples containing unique editing effects in the temporal domain. They are a fast-playing video and a time lapse video, respectively, which are interesting to viewers and thus ranked 1st and 2nd in MOS, respectively. However, as V-BLIINDS regards them as having poor motion consistency, it gives them undesirable low ranks (i.e., 40th and 44th, respectively).
\par In Section \ref{section4:metadata}, metadata were shown to be useful to extract quality information, showing better quality evaluation performance than the NR objective metrics. A limitation of metadata is that some are sensitive to popularity, users' preference, or other video-unrelated factors, which may not perfectly coincide with the perceived quality of videos. Video \#25 is such an example. It is a music video made by a fan of a musician. It has moderate visual quality and thus is ranked 25th in MOS. However, since the main content of this video is music, its popularity (the viewcount, the number of likes and comments, etc.) is mainly determined by the audio content rather than the visual quality. Moreover, it has the longest video description (listing the former works of the musician) in the dataset, according to which it would be ranked 14th (note that the description length is the best-performing metric in Table \ref{table_metasrocc}).
\par A way to alleviate the limitation of each metadata-driven metric is to combine several metadata-driven metrics and expect them to compensate for the limited information of each of them, which was shown to be effective in our results. For instance, the available maximum resolution was shown to be highly correlated with MOS in Table \ref{table_metasrocc}, so video \#29 would be ranked highly since a high-resolution (1080p) version of this video is available. However, it is ranked only 29th in MOS due to compression artifacts and packet loss artifacts. When multiple metadata-driven metrics are combined using SVR, the rank of the video becomes 24th, which is much closer to the MOS rank.


\section{Conclusion}
\label{section7}
We have presented our work on investigating the issue of the subjective and objective visual quality assessment of online user-generated video content. First, we examined users' patterns of quality evaluation of online user-generated videos via the HodgeRank decomposition and graph clustering techniques. A large amount of local inconsistency in the paired-comparison results was found by the HodgeRank analysis, which implies that it is difficult for human viewers to determine quality superiority between videos ranked similarly in MOS, mainly due to the difficulty of comparing quality across different factors. Consequently, subjective distinction between different quality levels is only clear at a large cluster level, which was shown by the graph clustering results. We then benchmarked the performance of existing state-of-the-art NR objective metrics, and explored the potential of metadata-driven metrics for the quality estimation of user-generated video content. It was shown that the existing NR metrics do not yield satisfactory performance, whereas metadata-driven metrics perform significantly better than the NR metrics. In particular, as each of the metadata-driven metrics covers only limited information on visual quality, combining them significantly improved the performance. Finally, based on the results and examples, we provided a detailed discussion on why the subjective and objective quality assessment of user-generated videos is difficult.
\par Our results demonstrated that the problem of quality assessment of user-generated videos is very different from the conventional video quality assessment problem dealt with in the prior work. At the same time, our results have significant implications for future research: The existence of diverse quality factors involved in user-generated videos and the failure of existing metrics may suggest that the problem of user-generated video quality assessment is too large to be conquered by a single metric; thus, metrics specialized to different factors should be applied separately (and then be combined later, if needed). Since many factors, such as editing effects, are not covered by existing metrics, developing reliable metrics for them would be necessary. Moreover, the problem is highly subjective and depends on personal taste, so personalized quality assessment may be an effective method in the future. Therefore, proper ways to collect personalized data as ground truths would be required, where big data analysis techniques may be helpful.
\par While our results are for a dataset with a limited number of video sequences, it is still reasonable to consider that most of them, particularly those related to the different nature of user-generated videos from professional ones, can be applied generally, although the matter of degree exists. Nevertheless, larger scale experiments with larger numbers of videos with more diverse characteristics will be desirable in the future.

%
%

\ifCLASSOPTIONcaptionsoff
  \newpage
\fi

\bibliographystyle{IEEEtran}
\bibliography{IEEEabrv,References_20150402}

\begin{IEEEbiography}[{\includegraphics[width=1in,height=1.25in,clip]{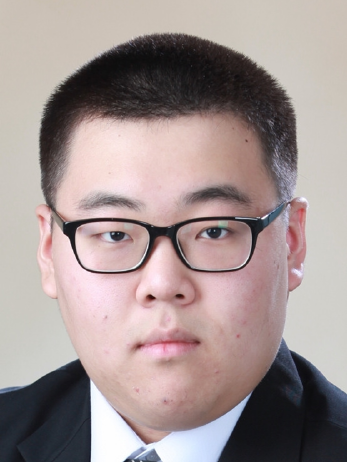}}]{Soobeom Jang}
 received the B.S. degree from the School of Integrated Technology, Yonsei University, Seoul, Korea, in 2014. He is currently pursuing the Ph.D. degree at the School of Integrated Technology, Yonsei University. His research interests include social multimedia analysis and multimedia applications.
\end{IEEEbiography}
\begin{IEEEbiography}[{\includegraphics[width=1in,height=1.25in,clip]{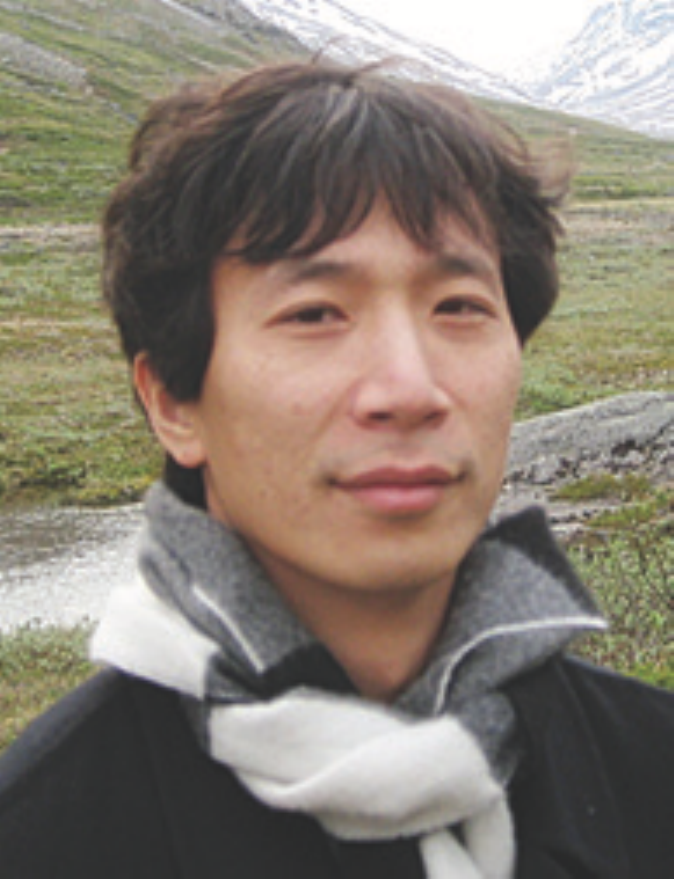}}]{Jong-Seok Lee}
 (M'06-SM'14) received his Ph.D. degree in electrical engineering and computer science in 2006 from KAIST, Korea, where he also worked as a postdoctoral researcher and an adjunct professor. From 2008 to 2011, he worked as a research scientist at Swiss Federal Institute of Technology in Lausanne (EPFL), Switzerland. Currently, he is an associate professor in the School of Integrated Technology at Yonsei University, Korea. His research interests include multimedia signal processing and machine learning. He is an author or co-author of about 100 publications. He serves as an Area Editor for the Signal Processing: Image Communication.
\end{IEEEbiography}

\end{document}